\documentclass[a4paper,fleqn,usenatbib]{mnras}
\usepackage{graphicx}	% Including figure files
\usepackage{amsmath}	% Advanced maths commands
\usepackage{mathtools}
\usepackage{amssymb}	% Extra maths symbols
\usepackage{comment,orcidlink}

\usepackage{threeparttable}
\usepackage{caption}
\def\msun{{\rm M_{\odot}}}

\def\be{\begin{equation}}
\def\ee{\end{equation}}

\def\del#1{{}}
\newcommand\mearth{{\,{\rm M}_{\oplus}}}
\newcommand\mj{{\,{\rm M}_{\rm J}}}

\usepackage[normalem]{ulem}

\newcommand{\SNc}[1]{\textcolor{red}{\textit{[\small #1]}}}
\newcommand\MSunPerYear{~${\rm M_{\odot}}$~yr$^{-1}$\,}

\title[Planets in tight binary systems]{Disc Fragmentation. III. The need for a new paradigm for formation of planets within close binary systems.}

\author[Luyao Zhang et al]{Luyao Zhang$^{1 \orcidlink{0009-0009-1163-2293}}$\thanks{lz239@leicester.ac.uk}, Sergei Nayakshin$^{1 \orcidlink{0000-0002-6166-2206}}$, Clement Baruteau$^2$, Philippe Thebault$^3$, Eduard I. Vorobyov$^{4,5}$\\
$^{1}$School of Physics and Astronomy, University of
  Leicester, Leicester, LE1 7RH, UK. \\
$^2$ IRAP, Universit\'e de Toulouse, CNRS, Toulouse, France\\
$^3$ LIRA, Observatoire de Paris, Université PSL, 5 Place Jules Janssen, 92195 Meudon, France\\
$^4$ Institut für Astro-und Teilchenphysik, Universität Innsbruck, Technikerstraße 25, 6020 Innsbruck, Austria,\\
}

\date{Accepted XXX. Received YYY; in original form ZZZ}

\pubyear{2026}

\begin{document}
\label{firstpage}
\pagerange{\pageref{firstpage}--\pageref{lastpage}}

\maketitle

\begin{abstract}
Dozens of planets and brown dwarfs are known to orbit one component of tight stellar binaries ($a_{\rm bin} \lesssim 20$ au), despite circumstellar discs in such systems being truncated to radii of only $\sim (0.2-5)$ au. This presents a challenge to classical planet formation models, which assume planets form after their host stars within stable discs. We propose instead that planet formation and binary formation are concurrent outcomes of gravitational fragmentation in massive circumstellar discs. In this scenario, rapid disc growth driven by infall from the parent molecular cloud leads to fragmentation at radii of tens of au, producing planetary-mass objects that migrate inward. Continued disc growth produces a dominant "oligarch" fragment that undergoes accretion runaway to become the secondary star. During this process, dynamical interactions eject many lower-mass planets, producing free-floating planets (FFPs), while others survive if they migrate sufficiently close to the primary star before destabilisation. Using numerical simulations, we show that survival depends strongly on formation time and mass. Planets formed early and those with masses $\gtrsim 1-3\mj$ are preferentially retained, whereas lower-mass planets ($\lesssim 0.1\,\mj$) are typically ejected. This mechanism naturally explains why low-mass planets are more deficient in tight binaries than gas giants, and predicts that FFPs have a steeper mass function than  bound planets within binaries.

\end{abstract}


\section{Introduction}\label{sec:introduction}

%The CA process is expected to operate only in binaries with wide separations and low eccentricities, where the planetesimal population remains dynamically cold \citep{Thebault_15_Stype_binaries}. The existence of these compact, massive, and dynamically excited systems therefore poses a major challenge to the CA scenario. Their high eccentricities and massive planets (even BDs) suggest that gravitational instability\citep[GI;][]{Boss97,HelledEtal13a,KratterL16} may play an essential role in their origin, beyond what standard CA can account for. 

The binary fraction of field FGK stars is about $\sim 50$\% \citep{RaghavanEtal10}. The binary separation distribution is quite broad, extending from as little as $\sim 0.01$~au to as large $\sim 10^4$~au, and shows a broad peak at $a_{\rm bin} \sim 40$~au \citep{Offner_23_binaries_review}. It is generally believed that wide binaries may form through a quasi-independent filament \citep[e.g.,][]{Inutsuka_Miyama_97_filament_fragmentation,Tomisaka_14_filament_fragmentation} or core fragmentation \citep{Inutsuka_92_core_fragmentation}. Close binaries (here defined as binaries with  separation $a_{\rm bin} < 100$~au) are more likely to form by disc fragmentation \citep{Offner_23_binaries_review}. This is supported by close binary fraction anti-correlation with host star metallicity \citep{Moe_Kratter_19_binary_vs_Z,Mazzola_20_binaries_vs_Z}, and by the fact that a number of multiple transiting stellar systems are now known \citep{Rappaport_22_triples,Kostov_24_triples}. These systems are well alligned and are hence likely to have formed from a common disc. Finally, disc fragmentation due to Gravitational Instability is expected to only occur at distances exceeding $\sim 50$~au \citep[e.g.,][]{Rafikov05,KratterL16}. Simulations of growing proto-binaries show that their separation may shrink as they mature \citep[][]{Artymowicz_91_binaries,BateB97,Tokovinin_Moe_20_binaries}, which could then explain naturally the location of the binary separation peak.

Binary stars provide us with a sensitive laboratory of planet formation theories. Observations show that binary properties influence planet formation process strongly. In contrast to the field binaries \citep{RaghavanEtal10}, the binary fraction of planet-hosting primary stars is significantly smaller, $\sim 22.5$\%, and the binary companion distribution peaks at hundreds of au \citep[although some of this could be due to adverse selection biases, see][]{Thebault_25_planets_in_binaries}. This comparison implies a detrimental effect of close binaries on planet formation \citep{Moe_21_planets_in_binaries}. 

There are two scenarios that may explain the deficit of planets in close binaries. Planet formation via Core Accretion \citep[CA;][]{PollackEtal96,IdaLin04a,Emsenhuber_23_CA} can be suppressed  due to high collision velocities of planetesimals  \citep[e.g.,][]{Heppenheim_78_binaries,Thebault_09_planet_in_binaries,PaardekooperEtal12, Marzari_13_CB_planet_formation, LinesEtal14}, the smaller initial disc mass budget and short gas depletion times in the binary-truncated circum-primary discs \citep[e.g.,][]{Thebault_15_Stype_binaries}.  
\citet{Venturini_26_planets_in_binaries,Nigioni_26_planets_in_binaries} recently presented a detailed population synthesis modelling of CA planet growth in circumprimary discs in binary systems. They include pebble accretion in their models, which speeds up planet growth. Despite this, \citet{Venturini_26_planets_in_binaries,Nigioni_26_planets_in_binaries} find a very strong reduction in planet formation in close binaries. In their models, for planets more massive than $10\,\mearth$, the limiting formation criterion is $a_{\rm bin} \gtrsim 33$~au, while giant planets with $M_{\rm p} > 100\,\mearth$ form only in binaries with $a_{\rm bin} > 40$~au. When perturbations from the secondary are taken into account, the induced increase in eccentricity further truncates the circumprimary disc and inhibits planet growth, thereby strengthening the constraints on planet formation \citep{Nigioni_26_planets_in_binaries}.

CA scenarions are therefore qualitatively consistent with the observed scarcity of planets in binaries with separation smaller than a few hundred au.  Observations, however, also revealed glaring exceptions to this expected suppression of planet formation. In the census of planet-hosting binaries compiled by \citet{Thebault_25_planets_in_binaries}, a number of systems have binary separations smaller than 30~au. A well-known example is the $\gamma$~Cephei system, which consists of stars with masses $M_1\sim 1.3\,\msun$ and $M_2\sim 0.33\,\msun$, separated by $a_{\rm bin}=20$~au, and with eccentricity $e_{\rm bin}=0.4$.  The primary hosts a gas giant planet of $6.6\,\mj$ at a separation of $a \sim 2$~au \citep{Chauvin_11_gamma_Ceph}. Although gas self-gravity may permit existence of dynamically quiet locations where planet embryo growth could still go on in binaries as tight as $a_{\rm bin}\sim 20$~au \citep[see][]{Silisbee_Rafikov_21_planets_in_binaries}, the additional formation challenges exposed by \cite{Venturini_26_planets_in_binaries,Nigioni_26_planets_in_binaries} challenge CA in this system strongly. 

HD~87646 is an eccentric ($e\sim 0.5$) binary with separation $\sim 20$~au \citep{Ma16,Fontanive_19_giants_in_wide_binaries}. Its circumprimary disc is expected to be cutoff at $\sim 3-4$~au (see Discussion). Yet the primary star hosts a $12.4\,\mathrm{M_J}$ planet at $0.12\,\mathrm{au}$ and a $57\,\mathrm{M_J}$ BD at $1.6\,\mathrm{au}$. Assumng that the disc that made these two massive objects was a few times their combines mass, we gather it must have been gravitationally unstable and be dispersed very rapidly. It is hard to see how either CA or the classic GI (see Discussion) could operate in such a small yet massive disc. 

Another remarkable system is HD~41004, a binary with separation of 23 au, in which both the primary and the secondary stars host S-type companions: a 2.5~$M_{\rm J}$ planet orbits the primary, and a 18~$M_{\rm J}$ BD orbits the secondary at a separation of less than 0.02~au \citep{Zucker03,Zucker04}. HD42936 (DMPP-3) currently holds the title of the tightest binary hosting an S-type planet: $a_{\rm b}\sim 1.2$~au, $e\approx 0.6$, with two small planet $M_{\rm p}\sim (1-3)\mearth$  planets orbiting the primary within $a\sim 0.06$~au \citep{Barnes_20_DMPP3,Stevenson_23_DMPP3,Standing_26_DMPP_planets}. In the standard scenario, the planets in the system would have to form within a disc extending to only $R\sim 0.2$~au. 

Observations of planets in binaries also show a confusing dependence of planet formation suppression on planet mass. Early on, \cite{Zucker02_giants_in_binaries} concluded that massive gas giants and BDs are found almost exclusively in binary stars. This is opposite to the trend for the planets in binaries as whole. Modern observations find that the global multiplicity rate of giant-planets hosts is higher than that of small-planet hosts \citep[e.g.,][]{FC_BG_21, MKU_MM_24}. These differences in multiplicity rates between giant and small planet hosts become especially pronounced when considering intermediate-separation binaries ($50<a<1000$ au) and close-in planets, for which the binarity rate of giant planet hosts is up to one order of magnitude higher than that for small planet hosts, and can even exceed that of field stars \citep{Ngo_16_Friends_of_HJ,Fontanive_19_giants_in_wide_binaries,ZieglerETAL_20,Moe_21_planets_in_binaries}. This strongly suggests that the formation process of small planets is much more hindered by binarity than that of giant ones, and that the formation of giant planets could even be favoured by intermediate-separation binarity\footnote{Even though multiplicity rates are affected by unavoidable observational biases \citep[see][]{Moe_21_planets_in_binaries,Thebault_25_planets_in_binaries}, these biases allone can probably not explain the very strong discrepencies that are observed between small and giant planet hosts \citep[see][]{FC_BG_21}}. These observational trends are counter-intuitive. Core Accretion formation of massive gas giants is far more demanding than that of low mass ones: the cores need to grow particularly massive to stage an accretion runaway, and this should occur whilst enough the gas is still present in the disc \citep[e.g.,][]{IdaLin04a,MordasiniEtal12}.

Disc fragmentation due to Gravitational Instability was postulated by \cite{Kuiper51b} to be capable of forming both planets and binary stars. Simulations in this century \citep{Gammie01,LodatoRice04,ZhuEtal12a} showed that discs can fragment only beyond $\sim 50$~au and on only relatively massive objects \citep{KratterEtal10,ForganRice13}. It is therefore widely believed that GI cannot form low mass (i.e., $M_{\rm p} \ll 1\mj$) and/or at short separation planets. However, due to rapid inward migration, and the role of dust concentrations collapse \citep[e.g.,][]{GibbonsEtal14,Longarini_23_solids_collapse_sims,Longarini_23_solids_collapse_theory,Rice_25_dust_collapse}, it is possible that GI results in formation of all types of planets \citep[see also][]{BoleyEtal10,ChaNayakshin11a,Nayakshin_Review}. GI remains a less well explored planet formation theory, and with many microphysical uncertainties, detailed predictions for planetary populations resulting from GI are very divergent \citep[][]{ForganRice13b,NayakshinFletcher15,MullerEtal18,Schib_25_dipsy_1,Schib_25_dipsy_2}, with some authors claiming no GI objects within $\sim 1$~au. However, the changes in the observed host star metallicity correlations for planets more massive than $\sim 5\mj$ and BDs \citep{Schlaufman18} do imply that most of these objects form via GI \citep[e.g.,][]{Nayakshin_Review}. Similarly, gas giant planets, $M_{\rm p} \gtrsim 0.1\mj$, orbiting M-dwarf stars are proposed to form via GI \citep[e.g.,][]{Schlecker_22_giants_in_Mdwarfs,Bryant_25_giant_around_Mdwarf}. Therefore, observations do provide some support for GI planets being able to migrate all the way inside 1 au.

%Numerical simulations show that likelihood of disc fragmentation due to GI increases when a massive companion is present in the disc \citep[][]{Meru15,Cadman_22_triggered_fragm_binary,Teasdale_25_trig_fragm}. This could potentially explain why the observed binary fraction for stars hosting massive giants and BDs is so high. Even in the absence of triggered fragmentation, the simple fact that formation of gas giants/BDs and the secondary star require a particularly massive disc may lead to a positive correlation in their frequency of appearance.

Motivated by the bewilderingly large abundance of Free-Floating Planets \citep[FFPs;][]{sumi2023,Mroz-23-Microlensing-planets-review,Yee-25-FFPs-MFunction} in papers I and II \citep{Calovic_25_FFP-1,Nayakshin_25_FFP-2}, we proposed that planets  in binaries may form by disc fragmentation {\em before} the secondary star itself is born. We found that in this scenario, as the secondary seed grows and migrates, planets are most often ejected, potentially contributing strongly to the observed FFP population. As a by product, simulations in papers I and II also showed examples of planets surviving in S-type and P-type (planets orbiting both binary stars) configurations. 
In this paper we explore this scenario further by performing numerical simulations designed to constrain the likelihood of planets surviving in the binary system rather than being ejected.

\section{The scenario}\label{sec:the_model}

We make the following assumptions in our model:
\begin{enumerate}
    \item An initially single protostar has formed as the result of the parent molecular cloud collapsing due to gravitational instability. The protostar is surrounded by a large and massive circumstellar disc that continues to gain mass from the cloud. The duration of this phase is comparable to the free-fall time of the parent cloud, $t_{\rm ff}\sim 0.1-0.3$~Myr \citep[e.g.,][]{Larson69}.
    
    \item As more and more distant regions of the cloud collapse, the angular momentum of the infalling material increases with time, and so does its centrifugal radius, $r_{\rm cf} = j^2 /(GM(R))$, where j is the specific angular momentum and $M(R)$ is the total enclosed mass within radius $R$. Due to the increase in $r_{\rm cf}$, the disc grows both in mass and in size \citep[cf.][]{VB06,VB10}, as long as the mass infall rate from the cloud onto the disk exceeds the mass transport rate through the disk onto the star.
    \item Disc fragmentation occurs at distances of tens to hundreds of au \citep{Rafikov05,Clarke09,LeeH_25_dusty_GI}, and results in a broad mass spectrum of gaseous fragments \citep{Xu_25_RHD_disc_fragmentation,Ni_25_GI_RHD,Vorobyov_25_clumps}. The duration of the disc fragmentation phase is comparable in duration to $t_{\rm ff}$.
    \item As previous simulations show, \citep[][also paper II]{Nayakshin17a}, fragments less massive than a few $\mj$ tend to migrate inward much more rapidly than they accrete gas. We neglect gas accretion on such fragments. These fragments are progenitors of planets.  
    \item Simulations show that the more massive the fragments are, the faster they accrete gas  \citep[e.g.,][]{Nayakshin17a}.     Fragments with mass $\gtrsim 10\mj$ accrete gas in a runaway regime, meaning that they accrete essentially all the gas supplied by the disc into their Hill sphere \citep[e.g.,][]{ZhuEtal12a}. Gas accretion for such fragments is taken into account in the simulations below; they are the progenitors of secondary stars born by disc fragmentation. We refer to them as "oligarch fragments" below. 
    \item Most fragments hatched by disc fragmentations are assumed to be in the low mass (planetary) regime. The oligarchs are more massive because they are born  at larger radii, where the Toomre mass is larger (see paper II), or they are the results of lower mass fragment mergers.
    \item As the result of these assumptions, the disc typically hosts a number of planets (including their remnants) by the time an oligarch is born.
\end{enumerate}

Putting these assumptions together, we propose that a typical initial condition for the formation of a close binary star is a single protostar growing rapidly by accretion of mass from a massive self-gravtating circumstellar disc. The system begins its transformation into a binary system when the disc hatches an oligarch fragment. Due to point (vi), and due to planet migration, many of the planets are likely to be on orbits interior to the oligarch, but exceptions to this can also take place. 

\begin{figure}
    \centering
    \includegraphics[width=0.5\textwidth]{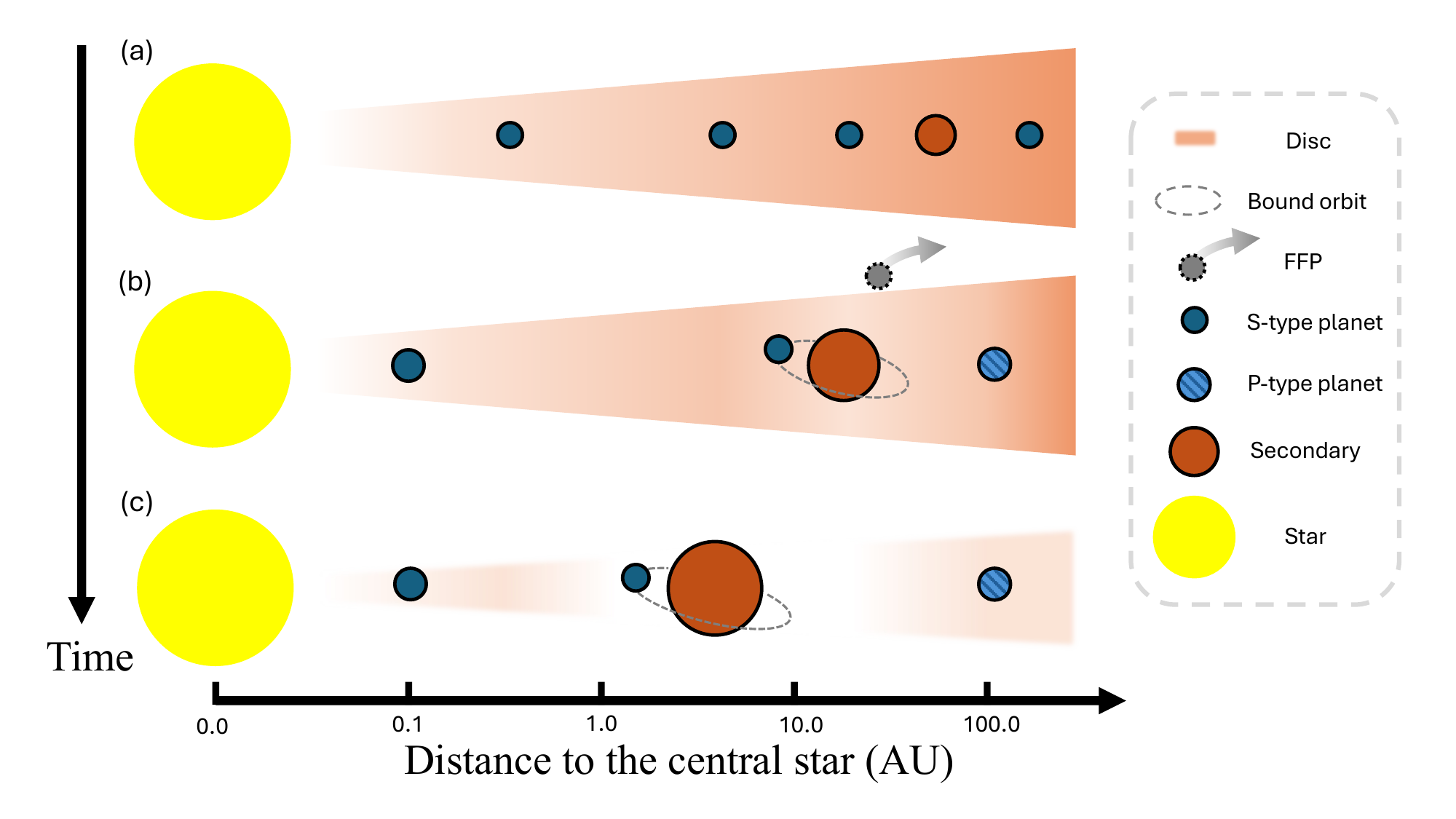}
    \caption{Schematic illustration of the model. (a) Several planets (the smaller blue circles) and the secondary seed ("oligarch", the larger brown circle) are born at $\sim 50-100$~au disc of the circum-primary disc. The innermost planet has migrated the deepest because it had been born first and/or it is more massive than the other planets. (b) The oligarch begins its rapid inward migration, while also gaining mass rapidly by gas accretion. It ejects one FFP and (may) have captured one of the planets on a bound orbit; (c) The final configuration of the system. The oligarch opens a deep gap/hole in the disc, and stops its rapid inward migration. The innermost planet survives as an S-type circumprimary planet. The outermost planet remains stranded at large radii as a wide orbit P-type planet. }
    \label{fig:sketch}
\end{figure}

The sketch in Fig. \ref{fig:sketch} illustrates this initial condition (panel a), and the sequence of events that we often observe in our simulations. The oligarch grows in mass and migrates in and may outpace some of the planetary mass objects. Eventually it opens a deep gap in the disc and its migration slows down strongly. Based on simulations performed in preparation of papers I and II, the following outcomes, in an increasing order of frequency of occurrence, are possible: (a) planets are ejected to become unbound FFPs; (b) others avoid ejection and become wide orbit circumbinary planets (as the P-type planet at $\sim 100$~au in Fig. \ref{fig:sketch}); (c) planets that migrated sufficiently close to the primary survive as S-type ones orbiting the primary; (d) very rarely, a planet is captured by the secondary, and becomes its S-type planet. 

\section{Numerical method}\label{sec:numerics}

As in paper II, 2D fixed grid hydrodynamic simulations are performed with \texttt{FARGO-ADSG}\footnote{\url{https://github.com/charango/dustyfargoadsg/}}, an extension of the original \texttt{FARGO} code by \cite{Masset00}, that includes gas self-gravity \citep{Baruteau_Masset_08_typeI} and an energy equation with radiative cooling are implemented \citep{Baruteau_Masset_08_typeI,Marzari_12_circumbinary_discs}. We set \cite{Shakura73} disc viscosity parameter $\alpha=0$ for our simulations. It should be kept in mind that GI generates its own turbulence, see \citep[][]{Gammie01,Rice05}. Shocks driven by GI cause a net heating and a net radial transport of angular momentum which are handled in the code by the use of a von-Neumann-Richtmyer artificial viscosity.  The ideal Equation of State (EOS) with a fixed adiabatic index $\gamma=1.4$ is assumed. Hydrodynamical equations are solved in the thin-disc 2D approximation, with radiative cooling given by radiation diffusion out of the disc. We set disc metallicity to Solar, and use \cite{Bell94} opacities.

Pre-collapse molecular H gas clumps that form by disc fragmentation are the progenitors of much denser post-collapse objects \citep{Larson69,Bodenheimer74,BodenheimerEtal80}. The former are resolved directly in simulations where the disc is massive enough to fragment (\S \ref{sec:fragmenting_disc}), whereas post-collapse planets and the secondary object are too compact to resolve via hydrodynamics. These are treated as sink/N-body particles with gravitational softening parameter $\varepsilon_s = 0.5 H$, where $H$ is the local disc scale-height. For N-body interactions, \texttt{FARGO-ADSG} uses the 5th order Runge-Kutta scheme, which we supplement by sub-cycling (cf. paper II) to increase the accuracy of the N-body integrator. The planets do not accrete gas in our models, given the low accretion efficiency discussed in \S~3 of Paper~II.  The secondary object accretion prescription is the same as in Paper~II, and was originally introduced by \cite{Kley_99_planet_accretion}. In this prescription, gas density within the secondary's Hill radius, $R_{\rm H}$, is reduced at a rate given by $f_{\rm acc} \Omega_{\rm s}$, where $\Omega_{\rm s}$ is the Keplerian angular velocity of the secondary around the primary, and $f_{\rm acc} = 0.1$. We find that the value of $f_{\rm acc}$ affects the secondary's mass from which the accretion runaway (cf. \S 3 in paper II) occurs, but it does not affect our main results or conclusions.

The central star mass is $M_*=1\msun$ in the simulations. In papers I and II, planets and the oligarch were injected into circular orbits at a range of locations in a gravito-turbulent but non-fragmenting disc. In \S \ref{sec:fragmenting_disc}, the disc mass is increased gradually until the disc fragments onto gas clumps. Such simulations allow us to probe formation of S-type planets in binaries in a setting in which the disc fragments ab-initio, but the outcomes are quite stochastic and hence we consider only a few examples of planets in tight binaries formed in such runs. In contrast, in \S \ref{sec:secondary_migration} we follow the approach of paper II, and study the case of a gravito-turbulent but non-fragmenting disc. We inject the seed of the secondary and the planets in this disc. These experiments are set in a more controlled fashion and enable us to start exploring the parameter space of the problem. For these reasons, \S\S \ref{sec:fragmenting_disc} \& \ref{sec:secondary_migration} employ different initial conditions, explained in the corresponding sections. Our radial grid is uniform in $\log R$, with open boundary conditions on both the inner, $R_{\rm in}$, and outer, $R_{\rm out}$, boundaries.

\section{A fragmenting disc example}\label{sec:fragmenting_disc}

\begin{table}
    \centering
    \begin{threeparttable}
    \caption{Initial and final parameters of collapsed objects in our simulation. }
    \label{tab:pla_par}
    \begin{tabular}{lcccc}
        \hline
        \textbf{Name} & \textbf{$t_\mathrm{inj}^\text{a}\,[\mathrm{kyr}]$} & \textbf{$a_0\,[\mathrm{au}]$} & \textbf{$m\,[\mj]$} & \textbf{$a^\text{c}_\mathrm{fin}\, [\mathrm{au}]$} \\
        \hline
        P1 & 2.1 & 24 & $1$  & 0.32 \\
        S1\textsuperscript{b} & 3.9 & 16 & $5-86$ & 13.5\\
        P2                   & 3.9 & 83  & $0.03$  & $\infty$  \\
        P3              & 3.9 & 158   & $0.01$  & 178.5\\
        P4              & 3.9 & 116   & $0.003$  & 478.9\\
        \hline
    \end{tabular}
    \begin{tablenotes}
        \small
        \item \textsuperscript{a} Injection time.
        \item \textsuperscript{b} S1 grows by accretion (cf. Fig. \ref{fig:RM_vs_t}).
        \item \textsuperscript{c} Final semi-major axis of surviving bound planets.
        % \item \textsuperscript{d} Velocity of ejected planets at infinity.
    \end{tablenotes}
    \end{threeparttable}
\end{table}

\begin{figure*}
	\centering
	\begin{minipage}{0.33\linewidth}
		\centering
		\includegraphics[width=1\linewidth]{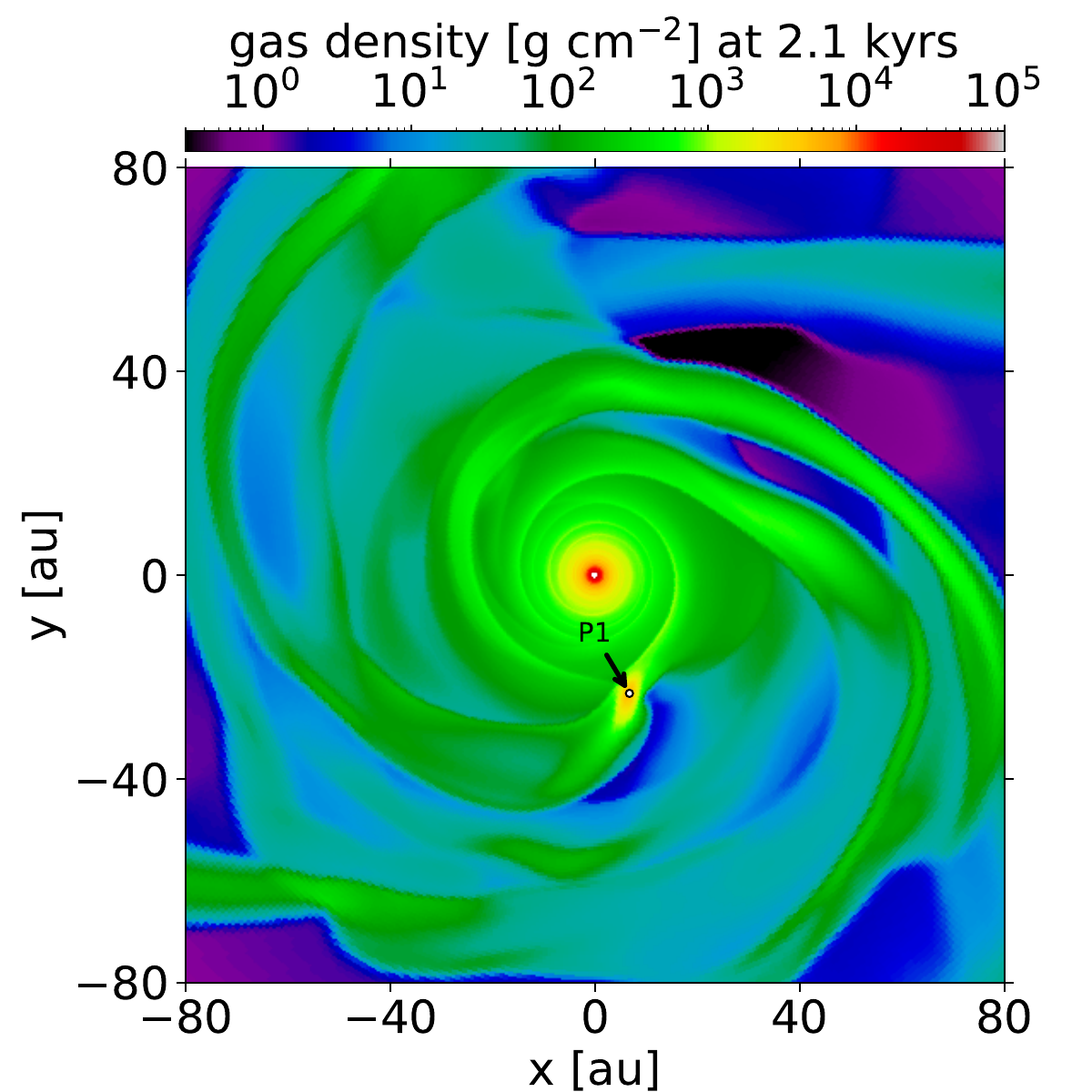}
	\end{minipage}
	\begin{minipage}{0.33\linewidth}
		\centering
		\includegraphics[width=1\linewidth]{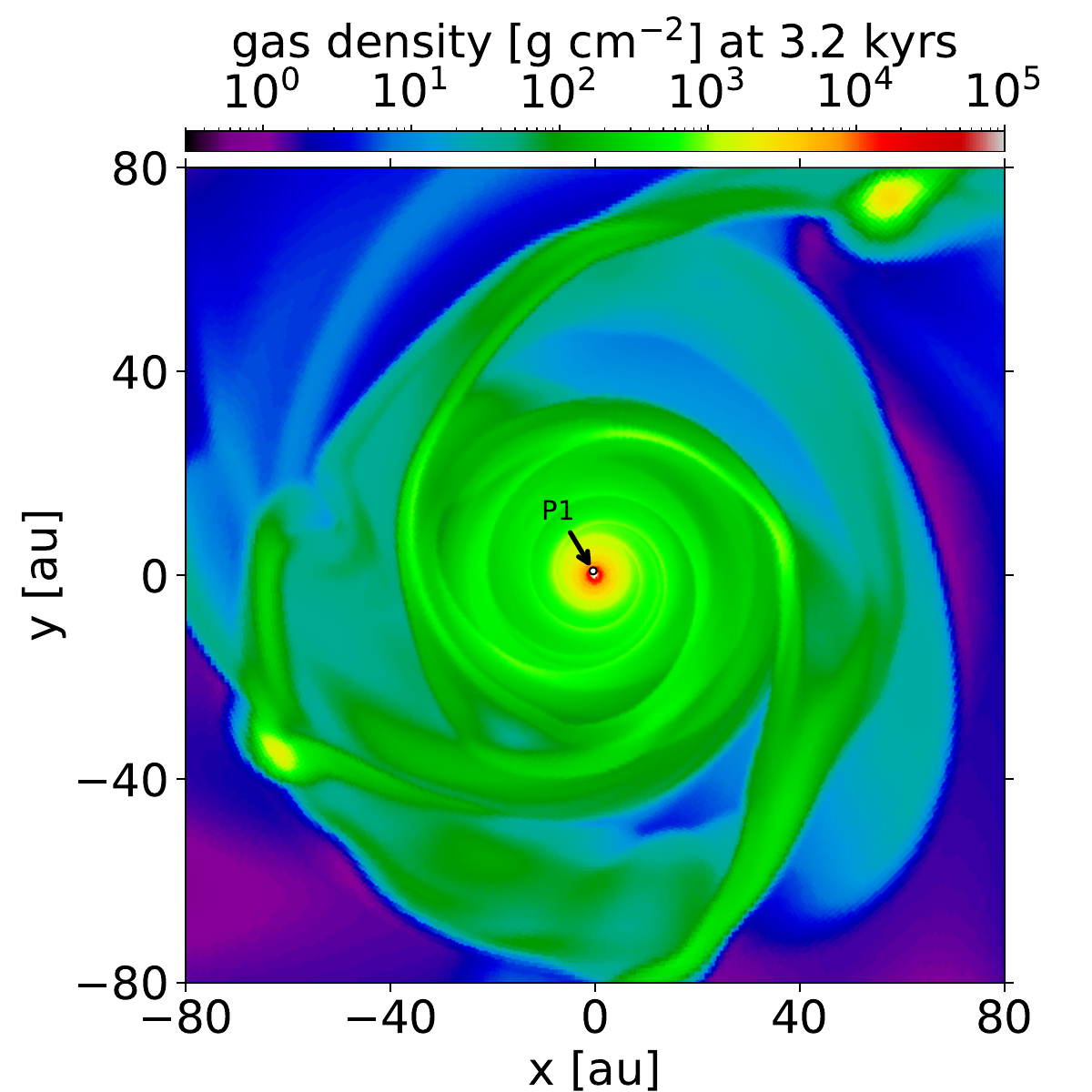}
	\end{minipage}
	\begin{minipage}{0.33\linewidth}
		\centering
		\includegraphics[width=1\linewidth]{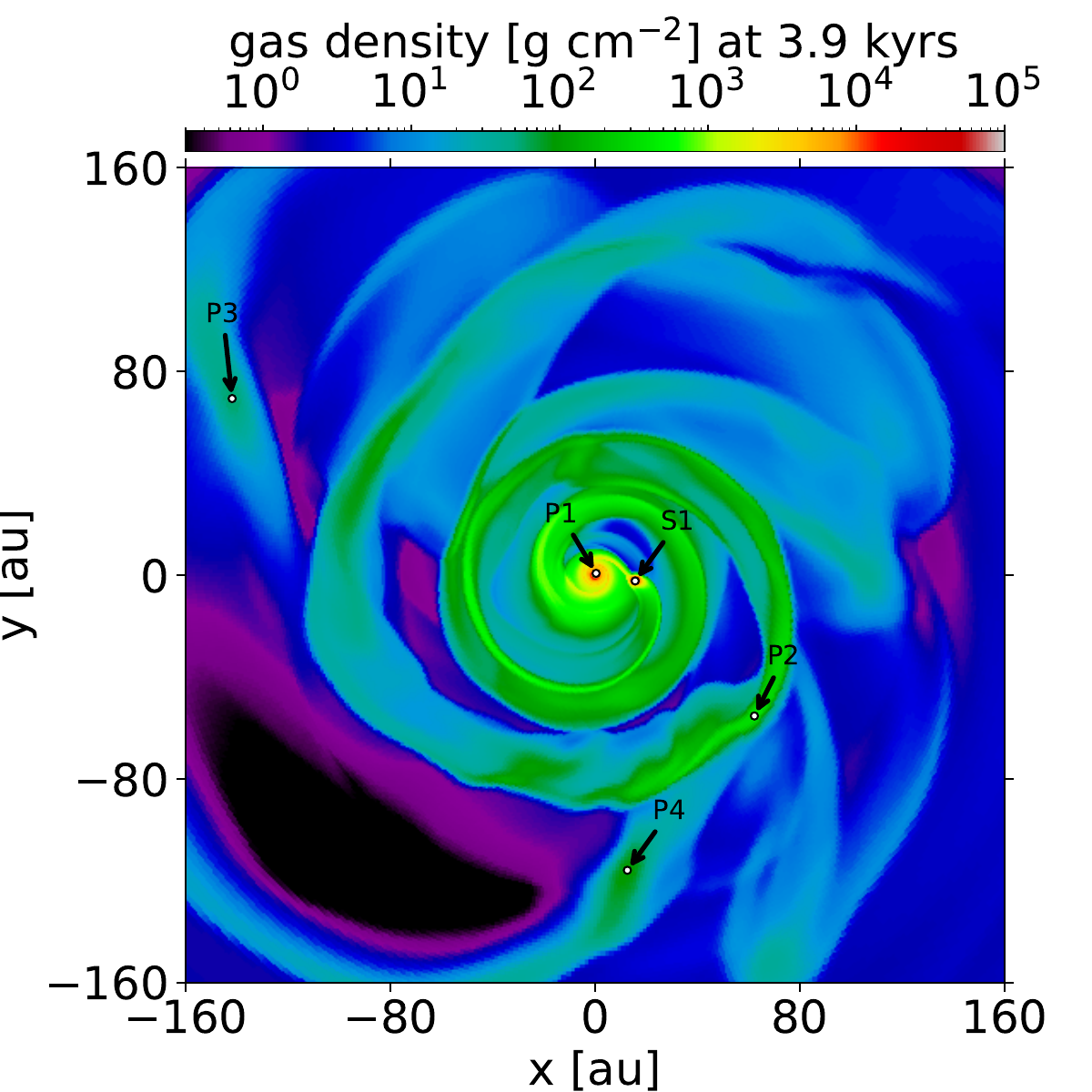}
	\end{minipage}
        \caption{The density for the disc at different times. White dots with black edges indicate the inserted objects. Panels: $t=2.1$~kyr, the insertion of planet P1 in the clump; $t=3.2$~kyr, P1 decouples from the disrupted clump and migrates inward;  $t=3.9$~kyr with zoom-out view, the secondary S1 is inserted into a massive and dense clump that migrated to $R\approx 16$~au, whereas low mass planets are inserted in dense filaments. }
    \label{fig:2D_disc}
\end{figure*}

In this section we simulate the process of disc fragmentation directly. Our radial grid  extends closer to the star than in paper II, down to $R_{\rm in}=0.5$~au. The outer radius is $R_{\rm out} = 1000$ au. The grid spans 700$\times$580 cells in the radial and azimuthal directions, making the grid cells approximately square at all radii. The grid resolution is $\Delta R/R \approx 0.011$. We estimate that the jeans length \citep[e.g., eq. 3 in][]{Vorobyov13c} is resolved by $\approx 10$ cells, which is sufficiently fine to capture disc fragmentation \citep[see also][]{ZhuEtal12a}. The initial surface density of disc follows $\Sigma(R)= \Sigma_0 (R_0/R)^{1.5} \exp[-R/R_{\rm cut}]$, with $\Sigma_0 = 2.5\times10^{-6} M_{\rm \odot}/(\mathrm{au})^2$, $R_{\rm cut}=150$ and $R_0 = 100$~au. The initial disc mass is $M_{\rm disc}\approx0.37 M_{\rm \odot}$. 

The disc is fed by an external mass deposition at a rate of $\dot M_{\rm dep}= 3\times 10^{-5}$\MSunPerYear into a narrow annulus at a radial distance of $R_{\rm dep}=100$~au. Unlike paper II, here we assume that the specific angular momentum of the incoming gas is lower than the local disc value, and is given by $j_{\rm dep} = 0.7 (G M_* R)^{1/2}$. This choice of $j_{\rm dep}$ smaller than the local Keplerian value mimics what is expected from collapse of rotating clouds. For example, \cite{Cassen_81_cloud_collapse} show that the angular momentum of the gas falling from the collapsing envelope is smaller than the local circular velocity value at the point where the stream strikes the protoplanetary disc. Additionally, 3D simulations of star formation in cluster environment often show that the direction of angular momentum of the gas falling into the disc from larger scales can fluctuate strongly \citep[e.g.,][]{Bate_19_star_formation_different_z}. The value of $j_{\rm dep}$ is relevant to the exact location and timing of disc fragmentation, but not our main results or conclusions.

While our simulations have sufficient resolution to capture disc fragmentation, yet higher resolution modelling of fragment centres is needed to capture clump collapse by H$_2$ molecule dissociation \citep[e.g.,][]{Bodenheimer74,Bodenheimer78}. Such a collapse forms progenitors of Jupiter-density gas giant planets or more massive objects. On the other hand, smaller, solid core dominated planets may form by grain growth and sedimentation inside gas fragments before H$_2$ dissociation \citep[e.g.,][]{VE19,HelledEtal08,ChaNayakshin11a}. Both processes are key to the evolution of fragmenting discs. Unfortunately, our simulations include neither a dust component nor H$_2$ dissociation, and  have insufficient resolution to model these processes. As the first step towards improving self-consistency of our simulations, in this section we identify the densest fragments that form by disc fragmentation, and inject post-collapse objects in their centres, assigning them the local gas velocity.

We performed a few such simulations aimed at exploring  formation of S-type planets specifically. While clumps as massive as low mass stars form in our simulations frequently, most result in binary separation of a few tens au, consistent with the observed binary statistics \citep{RaghavanEtal10}. For further analysis in this section, we selected the simulation that resulted in the tightest binary and an S-type planet around the primary.

As the disc mass grows due to external mass deposition, it fragments at $R\sim 70$~au into half a dozen clumps with initial masses of $\sim 5$--$10\,\mj$ at $\sim 1.7\times 10^{3}$ years, when the disc mass reaches $\sim0.42 M_{\rm \odot}$.  Most of these clumps are disrupted locally, often in about one orbit; others merge, and a few migrate inward before being destroyed by stellar tidal forces. We selected a particularly sturdy clump that was able to migrate deep into the inner disc to study what would happen if its centre did collapse to form a gas giant planet. We interrupt the simulation just before the clump is destroyed, when its mass is $M\approx 6.7\,\mj$, and it is located at $R\approx 24$~au from the star. We inject a planet of mass  $M_{\rm p}=1\,\mj$ (P1 in Table~\ref{tab:pla_par}, treated as a gravitationally softened point mass without gas accretion), into the centre of the clump, and restart the simulation. In Appendix A we vary the mass of P1 to explore the sensitivity of our model to this parameter.

The left panel of Fig. \ref{fig:2D_disc} shows the density for the disc at that moment. As in the original simulation, the fragment hosting P1 is tidally shredded quickly after restart, but P1 continues to migrate inward. At the same time, fragments continue to form in the outer disc (e.g., see the middle panel of Fig. \ref{fig:2D_disc}). Eventually, a more massive gas clump forms and migrates into the inner disc. At $t = 3.9$~kyr (third panel of Fig.~\ref{fig:2D_disc}), the clump has migrated to $R \approx 16$~au and grown to $M \approx 45\,M_{\rm J}$, with a central temperature of $\sim 1400$~K. It is quite likely that such a clump would undergo second collapse by H$_2$ dissociation (which we do not resolve here). Therefore, we interrupt the simulation once again and insert a seed secondary of mass $M_{\rm s} = 5\,M_{\rm J}$ \citep[][corresponding to the typical mass of a protostellar seed formed after the second collapse triggered by H$_2$ dissociation]{Vaytet+13, Bhandare+19, Ahmad23} at the clump centre. In Appendix \ref{apd:test}  we vary the initial mass of S1 to explore the sensitivity of our model to this parameter.

\cite{GibbonsEtal14,Longarini_23_solids_collapse_theory,Longarini_23_solids_collapse_sims,Rice_25_dust_collapse,Baehr_2022,Rowther_2024} find that grains focused inside spiral density arms of gravito-turbulent discs can collapse by direct gravitational collapse into $M_{\rm p} \sim 0.1-10\mearth$ solid cores. In paper II we injected such cores into the disc at arbitary locations. Here we injected three $10\mearth$ cores into the densest regions of spiral density arms (one core per spiral arm) at the time when the secondary S1 is introduced. In the right panel of Fig.~\ref{fig:2D_disc}, we show all the solid cores, marked P2, P3 and P4 in Table \ref{tab:pla_par}, along with the Jupiter mass planet (P1) and the secondary (S1).

\begin{figure*}
    \centering
    \includegraphics[width=0.45\textwidth]{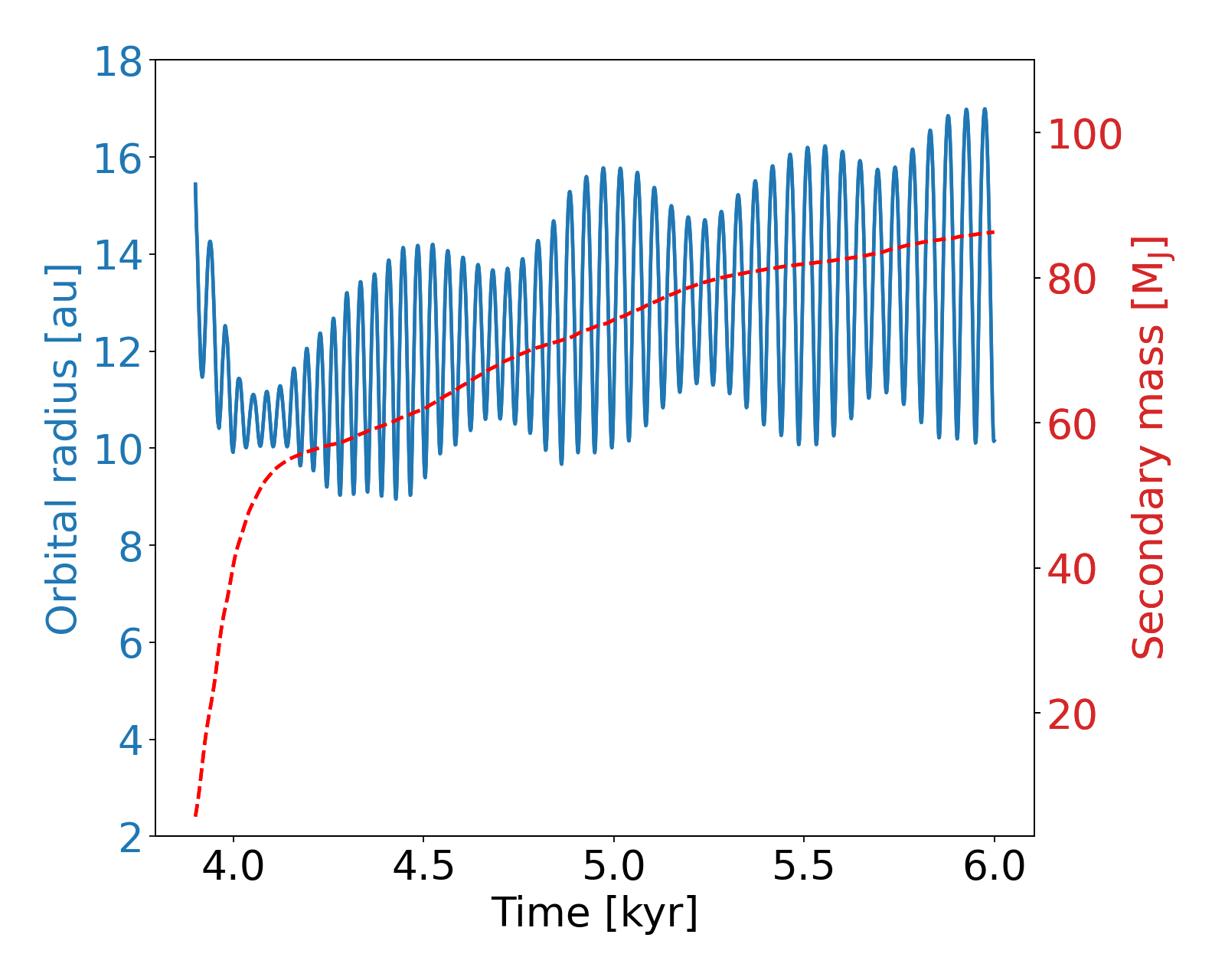}
    \includegraphics[width=0.45\textwidth]{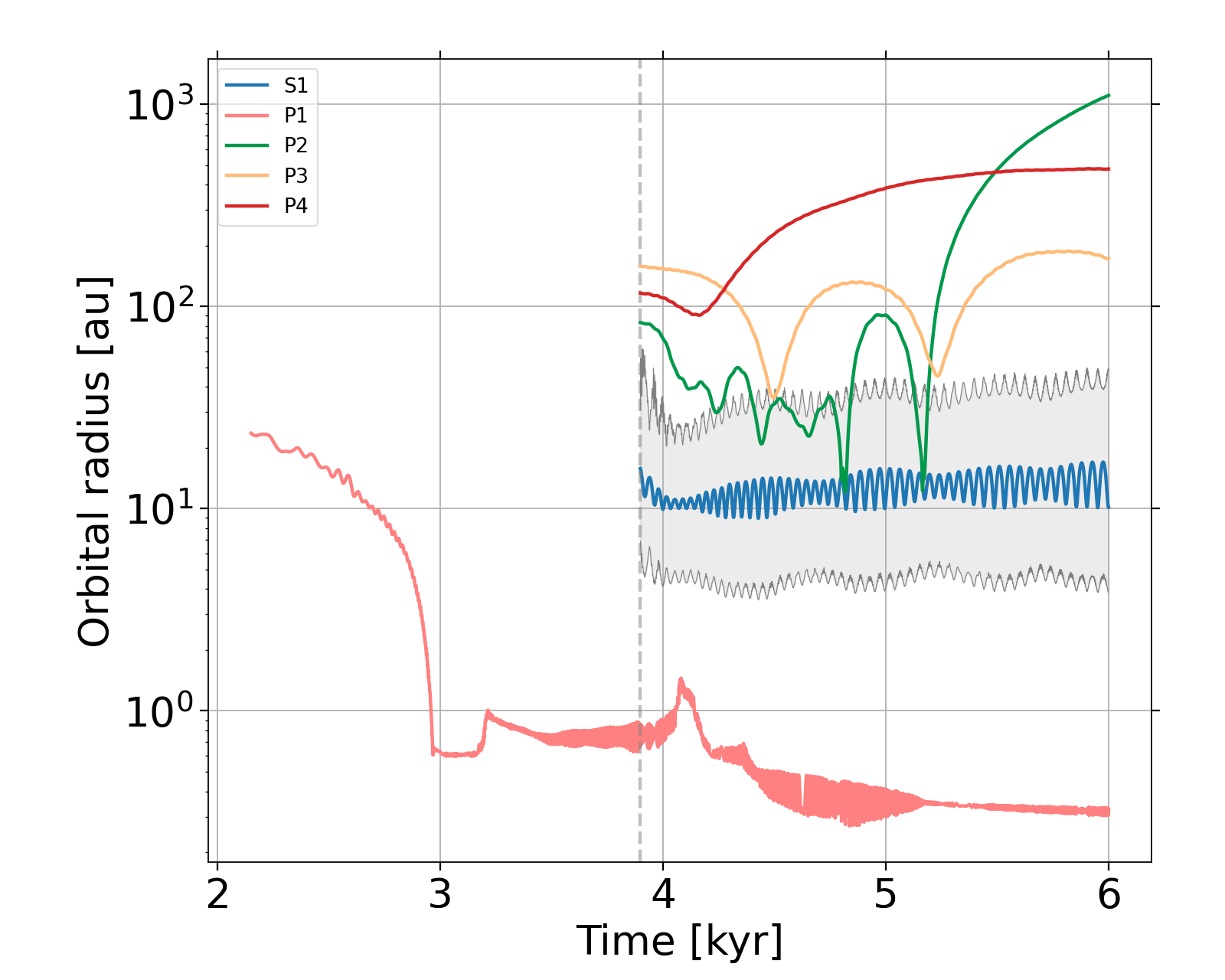}
    \caption{Left: Time evolution of the secondary's orbital radius and mass. The left y-axis shows the orbital radius (blue curve), while the right y-axis shows the secondary mass (red curve).
    Right: Orbital radii of the secondary and planets in the simulation. The grey shaded region indicates the radially unstable range according to the criterion of \citet{Holman_Wiegert_99}. The grey dotted line marks $t=3.9\,\mathrm{kyr}$, when the secondary and the other planets were injected. }
    \label{fig:RM_vs_t}
\end{figure*}

\begin{figure}
    \centering
    \includegraphics[width=0.45\textwidth]{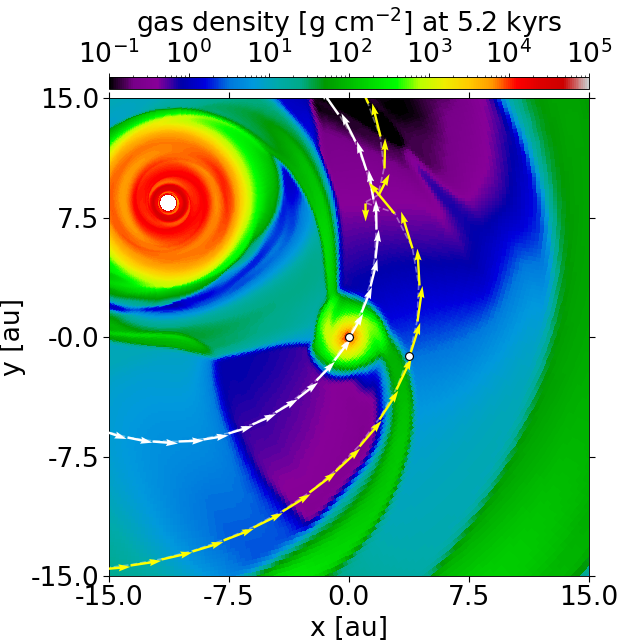}
    \caption{Surface density field of the disc at $t = 5.2\,\mathrm{kyr}$. Note that the figure is centred on the secondary. The primary star is towards the top left corner of the figure, inside the (white) inner boundary region. The white dashed lines trace the trajectories of S1 and P2 in the frame fixed on the primary, while arrows indicate their velocities. The separation of P2 and S1 at the closest approach is $0.6\,\mathrm{au}$.}

    \label{fig:P2_kicked}
\end{figure}

The simulation is then restarted again. The fragment hosting S1 is quickly consumed by the secondary seed as it grows in mass, whereas the spiral arms hosting the embedded cores dissipate away. Then, following its interaction with S1, P2 is ejected. P3 evolves more smoothly through a more distant encounter with S1 that lifts its semi-major axis but does not eject fully. Core P4 is pushed outward by the spiral structure rather than the secondary; this is consistent with the results of \cite{Rowther_Meru_2020}, who showed that low mass planets in self-gravitating discs with realistic cooling prescription (rather than constant $\beta$-cooling) may migrate outward. 

Fig. \ref{fig:RM_vs_t} presents evolution of the secondary seed's separation and mass (left panel), and of the separation for all of the objects from Table \ref{tab:pla_par} (right panel). Comparing the tracks of P1 and S1, we observe several key differences. P1, with its lower mass, remains in the type I migration regime, and is able to migrate nearly all the way to the inner grid boundary, $R_{\rm in}$. It then hangs at $R\approx 0.6$~au for a time. This temporary stalling of P1 is an artifact of a finite value of $R_{\rm in}$ and our open boundary conditions that create a strong positive gradient in the disc surface density there, which artificially traps P1 there \citep[cf.][]{Masset06_planet_traps}. We find that such a balance is not always stable, however, since P1 is rather massive and may affect the disc structure itself (this is further discussed in \S \ref{sec:single_object}). Due to this, P1 undergoes two abrupt orbital jumps at $t\approx 3.2$ and $4.1$~kyr, and finally tumbles inside the inner boundary at $t\approx 4.5$~kyr. The details of P1 migration into the region $R\lesssim 0.5$~au are affected by the numerical artefacts inside the inner boundary $R_{\rm in}$. Since this region of the disc is not simulated, the inner wake of the planet is unresolved, which implies that P1's migration cannot be handled well by the code at that point. However, previous 1D disc calculations show that gas giants can migrate as far in as 0.1 au in very young, rapidly accreting discs \citep[e.g.,][]{Nayakshin-23-FUOR}. 

In contrast, S1 becomes very massive rapidly, consuming the parent clump and then opening a deep gap in the disc. At that point it switches from the rapid type I to the slower type II migration. The disc displays significant eccentricity and fluctuating spiral density arms, which strongly affect the dynamics of S1 in turn. At the end of the simulation, S1 is on an eccentric orbit with $a\approx 14$~au and $e\approx 0.3$.

The grey shaded area in the right panel of Fig. \ref{fig:RM_vs_t} shows the radial range within which planetary orbits are unstable to ejection according to the \cite{Holman_Wiegert_99} criteria. In accordance with that, planet P2 is ejected, whereas P1 survives as an S-type planet bound to the primary star, and P3, P4 survive in wide orbits.  Fig.~\ref{fig:P2_kicked} shows the disc density structure at $t = 5.2\,\mathrm{kyr}$, shortly before the last close P2-S1 interaction. The secondary, surrounded by a circumstellar disc, is located at the centre of the figure. The white dashed curves show the orbits of S1 and P2 close to the time shown in the figure, with the arrows depicting their velocities. The primary-centric velocity of P2 is $\approx 11$ and $14\,\mathrm{km\,s^{-1}}$ just before and just after the close encounter, respectively, indicating that P2 received a kick of $\sim 3 \mathrm{km\,s^{-1}}$ due to its interaction with S1. Fig.~\ref{fig:P2_kicked} shows  that during the encounter the planet moves retrograde for a period of time. The velocity of P2 at infinity is $\sim 6$~km/s.

\section{Non fragmenting disc}\label{sec:secondary_migration}

\subsection{A single embedded object}\label{sec:single_object}

In this section we present examples of a single fixed mass object evolving in a massive gravito-turbulent disc. Such simulations are less self-consisent than those from \S \ref{sec:fragmenting_disc}, but they are useful in exploring the parameter space of the problem since they are set in a more controlled way. 

For these experiments, the computational grid extends from $R_{\rm in}=1$~au to $R_{\rm out}=10^3$~au, and the approximately square cell grid has 560$\times$500 cells in the radial and azimuthal directions, resulting in grid resolution $\Delta R/R \approx 0.0124$. 

\begin{figure}
    \centering
    \includegraphics[width=0.45\textwidth]{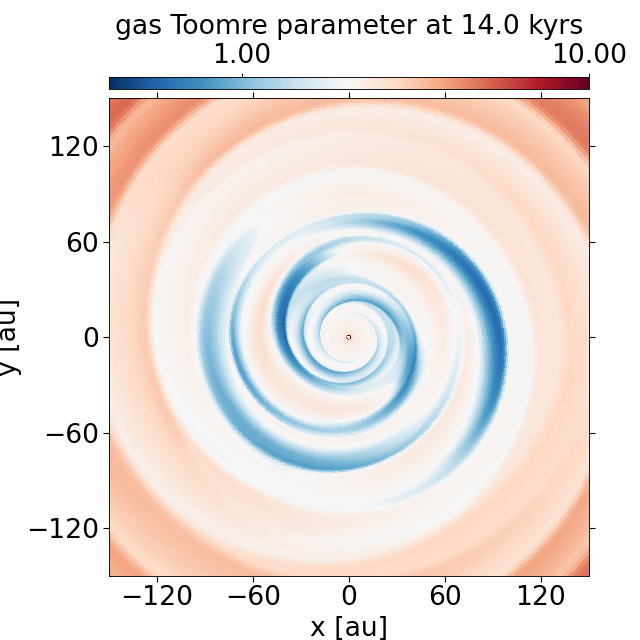}
    \caption{The Toomre $Q$-parameter field for the relaxed disc at $t = 14\,\mathrm{kyr}$, shown using a red--blue divergent colour scheme with a transition at $Q \approx 1.4$.}
    \label{fig:relax_disc}
\end{figure}

The gravito-turbulent but non fragmenting disc initial condition is obtained as following. At time $t=0$, the surface density of the disc is given by $\Sigma(R)= \Sigma_0 (R_0/R)^{1.5} \exp[-R/R_{\rm cut}]$, with $\Sigma_0 = 5\times10^{-5} M_{\rm \odot}/(\mathrm{au})^2$, $R_{\rm cut}=100$ and $R_0 = 100$~au. The disc is fed by an external mass deposition at a rate of $\dot M_{\rm dep}= 1\times 10^{-5}$\MSunPerYear into a narrow annulus at a radial distance of $R_{\rm dep}=50$~au \citep[compared with \S \ref{sec:fragmenting_disc}, the smaller values of $\dot M_{\rm dep}$ and $R_{\rm dep}$ here preclude disc fragmentation, see, e.g.,][]{Clarke09,ZhuEtal12a}. At time $t=5 \times 10^3$ years, the external mass deposition is turned off. We continue the simulations until $t=1.4\times 10^4$ years, allowing the disc to relax further. The resulting \cite{Toomre64} $Q$-parameter is plotted in Fig. \ref{fig:relax_disc}. The region from tens to $\sim 100$~au has Toomre parameter of order unity, so we expect that the disc could hatch planetary mass bodies in that region.

\begin{figure*}
    \centering
    \includegraphics[width=0.45\textwidth]{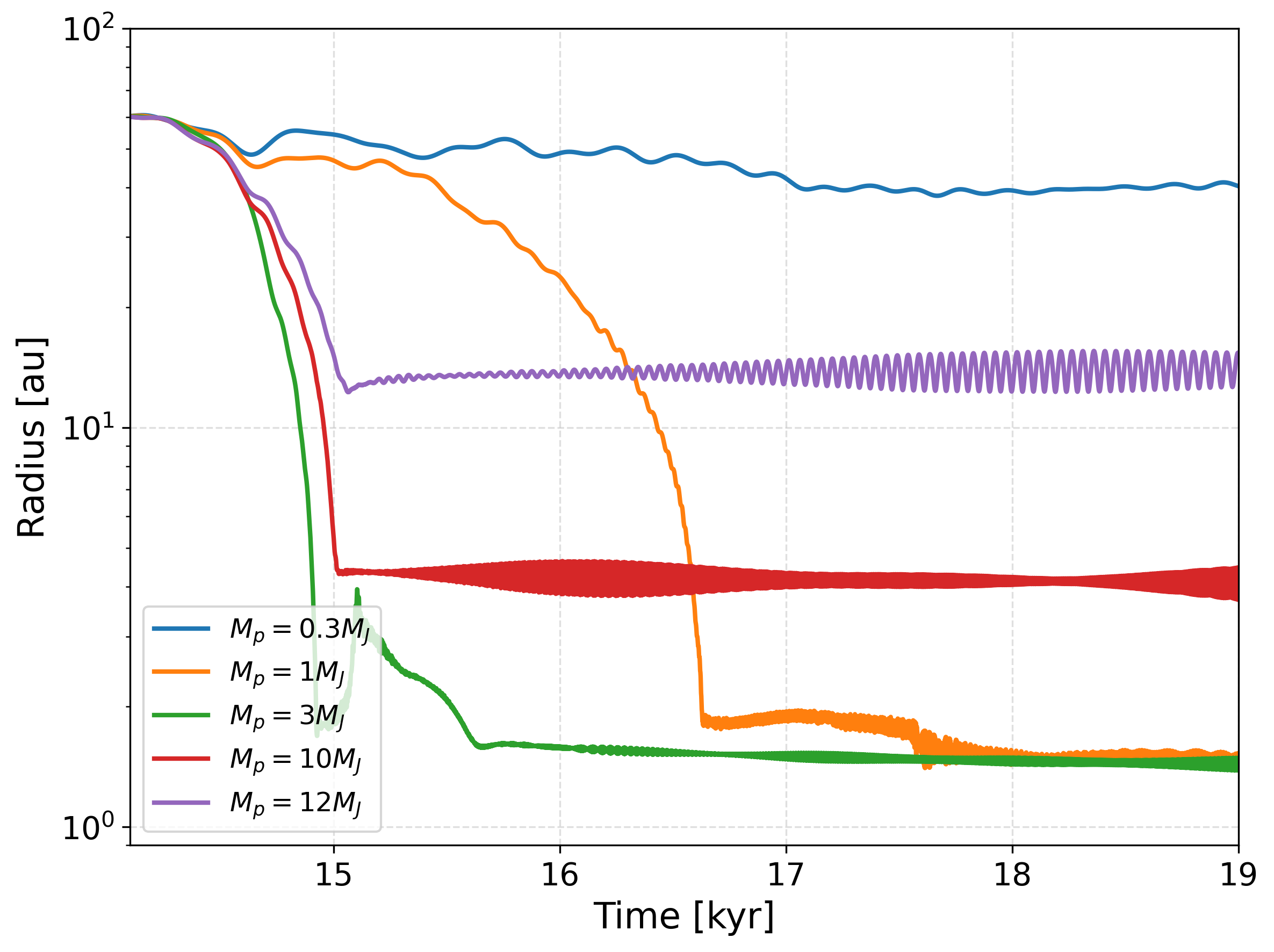}
    \includegraphics[width=0.45\textwidth]{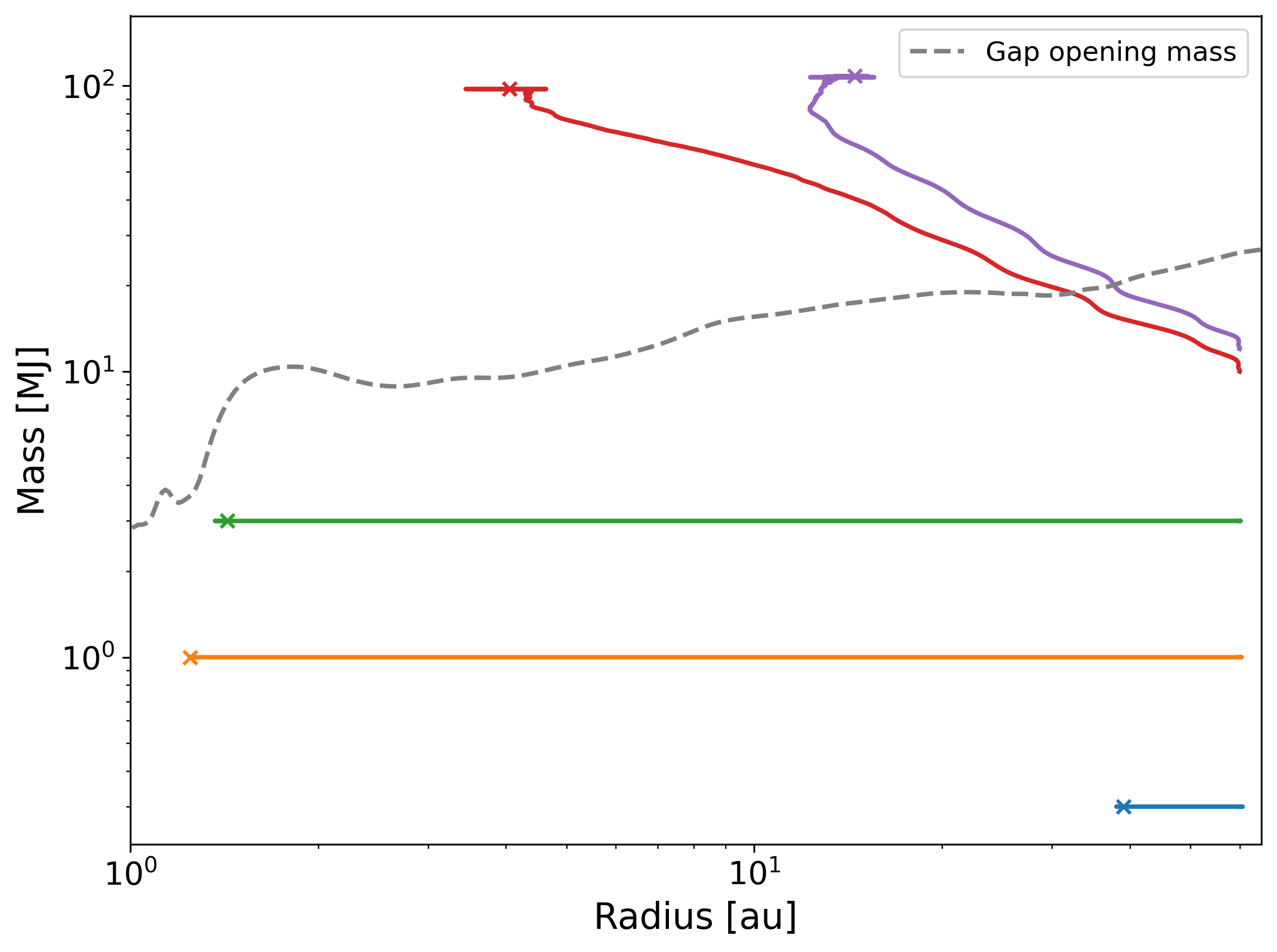}
    \caption{
    Left: Evolution of the orbital radius for the secondary object with masses $M_p = 0.3, 1, 3, 10, 12\,M_{\rm J}$. 
    Right: Evolution of the object mass and radius in the mass-radius plane for the simulations shown on the left. The crosses on the solid curves mark the time of 19 kyr. The dashed curve shows the gap-opening mass, $M_{\rm gap}$, for the disc at the time of object injection into the disc. Tracks of objects $M_p = 0.3, 1, 3\mj$ are horizontal since gas accretion is disallowed for them (cf. \S 3 in paper II). For massive objects, $M_p = 10, 12\,M_{\rm J}$, gas accretion is allowed. Note that soon after they cross the gap opening mass curve, their tracks become mainly vertical -- they accrete gas more rapidly than they migrate.}
    \label{fig:sec-motion}
\end{figure*}

In this section, one object per disc is injected at  $R=60$~au into the disc. The left panel of Fig. \ref{fig:sec-motion} shows the radius versus time evolution for objects with a range of initial masses. In accordance with our model, for the low mass objects, $M_{\rm p} = 0.3, 1$ and $3\mj$, gas accretion is disallowed. For the two more massive ones, $M_{\rm p} = 10$ and $12\mj$, accretion is included as described in \S \ref{sec:numerics}, and they grow to mass $M_{\rm s} \sim 0.1 \msun$ rapidly. In the right panel of Fig. \ref{fig:sec-motion}, we show the tracks of the two massive objects in the mass-separation plane, along with those of the non-accreeting planets.

Jupiter mass objects do not open gaps in massive self-gravitating discs \citep{BaruteauEtal11,MalikEtal15}, and therefore they migrate in a type-I like \citep{CridaEtal06} regime.  In this regime, the migration timescale scales approximately as $t_{\rm mig}\propto M_{\rm p}^{-1}$. This explains why $M_{\rm p}=3\mj$ planet migrates most rapidly in the left panel of Fig. \ref{fig:sec-motion}. Interestingly, $M_{\rm p} = 0.3 \mj$ planet (or any of the less massive planets that we experimented with), did not manage to migrate into the inner disc at all, even by as late in time as $t = 50$~kyr (not shown in the figure), and some even migrated outward. We believe that this is caused by the strong stochastic gravitational torques driven by the spiral density arms \citep{Rowther_Meru_2020}, analogous to the type-I migration of low mass planets in turbulent non self-gravitating discs \citep{CB_Lin_2010,Wu_Yinhao_24_stochastic_migration}. We note that in other experiments we conducted, low mass (i.e., $M_{\rm p}\lesssim 0.3\mj$) gas giants were often able to migrate to the inner boundary, but it always took them substantially longer than it did for super-Jupiter planets \citep[see also][]{BaruteauEtal11}.

In the right panel of Fig.~\ref{fig:sec-motion}, the dashed line shows the gap opening mass, $M_{\rm gap}$, for the disc at the time of planet injection. At this planet mass, the \cite{CridaEtal06} parameter is equal to 1, and the planet is expected to open a very deep gap. $M_{\rm gap}$ is computed using eq. 10 from \cite{BaruteauEtal14a}, assuming the \cite{Shakura73}  disc viscosity parameter of $\alpha = 0.03$, as relevant for gravito-turbulent discs \citep[e.g.,][]{Gammie01,Rice05,BaruteauEtal11}.  Our two "oligarch" mass objects initially migrate inward very rapidly in the type-I regime, crossing the $M_{\rm gap}$ line at $30-40$~au, when they reach the mass of $\sim 25\mj$. Since gap opening takes a finite time \citep{MalikEtal15}, they continue to migrate in some more, gaining more and more mass. Eventually, the oligarchs start migrating in the much slower type II regime.

\begin{figure*}
    \centering
    \includegraphics[width=0.49\textwidth]{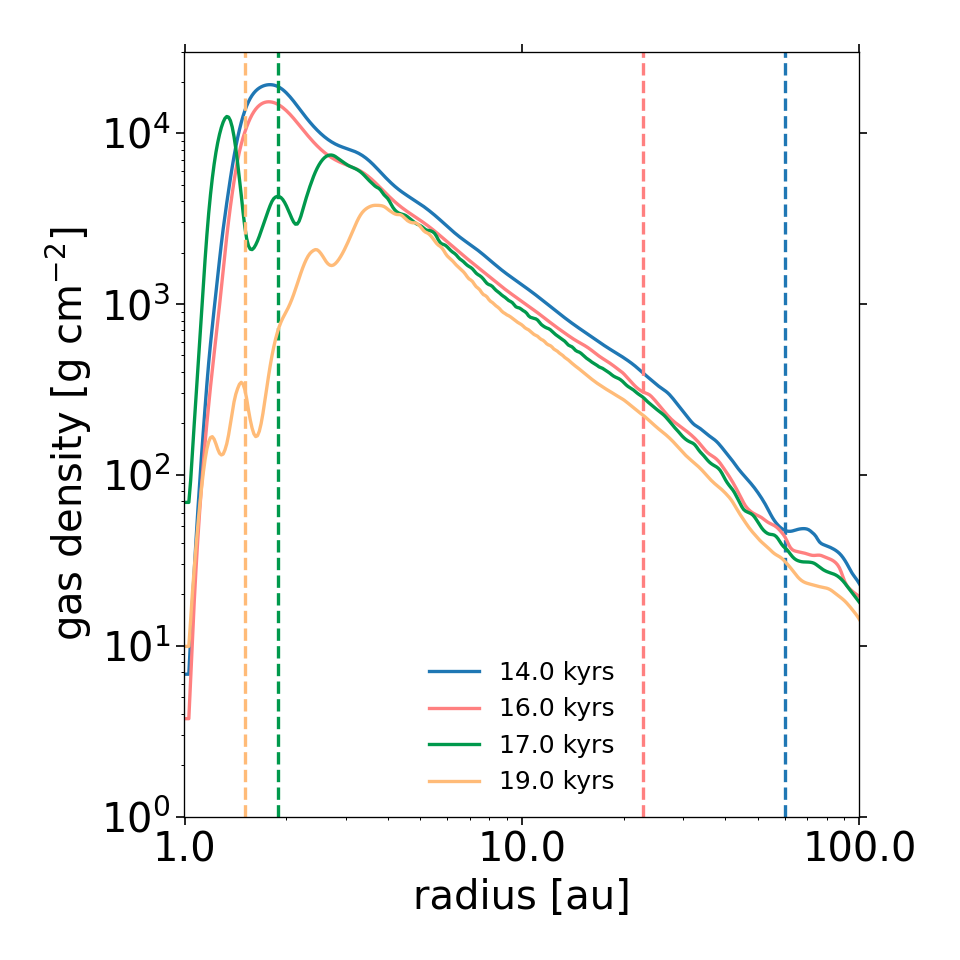}
    \includegraphics[width=0.49\textwidth]{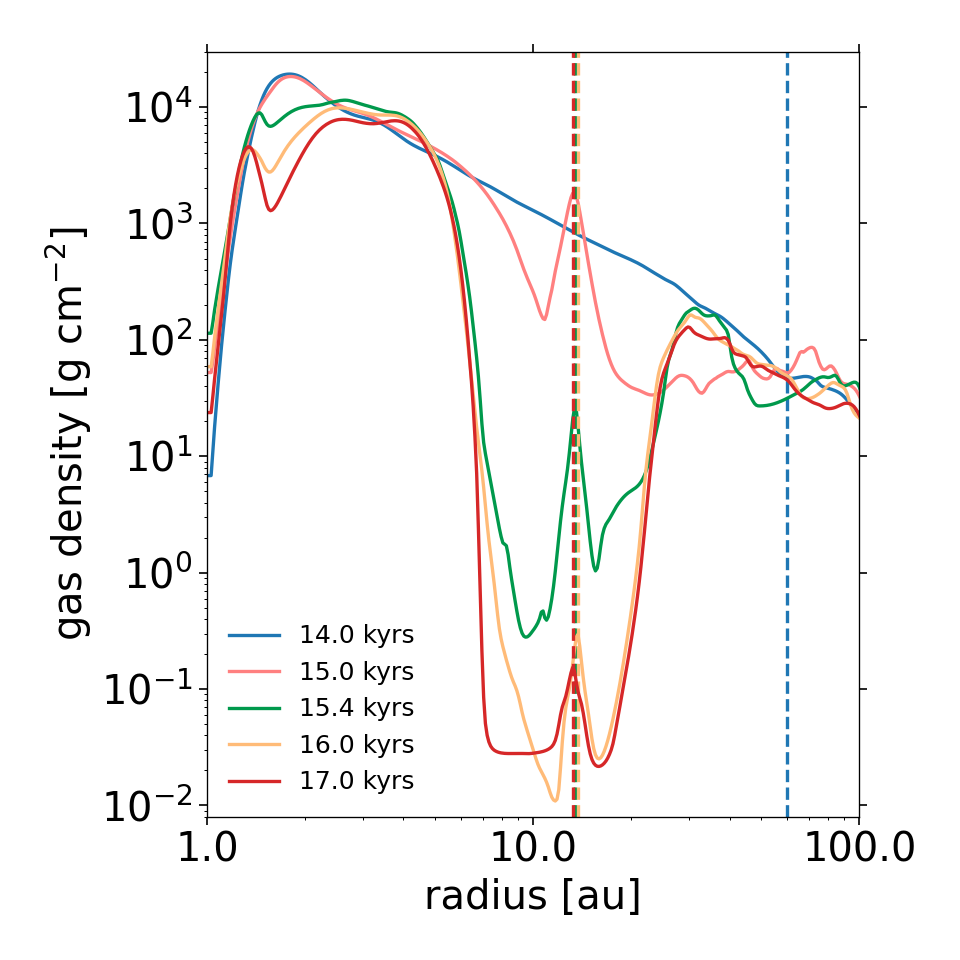}
    \caption{The azimuthally averaged disc surface density for simulations with $M_{\rm p} = 1\,\mj$ at $t = 14,16,17,19\,\mathrm{kyr}$ (left panel) and $M_{\rm p} = 12\,\mj$ at $t = 14,15,15.4,16,17\,\mathrm{kyr}$  (right panel). Dashed vertical lines indicate the planet position at the corresponding times.}
    \label{fig:1d-disc}
\end{figure*}

Note that the $M_{\rm p} = 1, 3\mj$ planets migrate to $R\sim 2$~au but may migrate outward and then back, similar to planet P1 in \S \ref{sec:fragmenting_disc}. As we noted there, the planet stops close to the inner boundary of our computational domain due to a strong positive gradient in $\Sigma$. This is an artefact of the open boundary conditions. We also suggested in \S \ref{sec:fragmenting_disc} that the back-and-forth motion of the planet occurs due to the sculpting of the disc near the boundary by the planet; its mass is smaller but comparable to $M_{\rm gap}$ at $R_{\rm in}$. 
Fig. \ref{fig:1d-disc} shows the azimuthally averaged disc surface density for the $M_{\rm p} = 1\mj$ and $M_{\rm p} = 12\mj$ simulation for several representative times. For the $M_{\rm p} = 1\mj$ case, we observe that before the planet reached its final position ($\sim 1.7$~au), the disc surface density there is as high as $10^4$g/cm$^2$. However, as the planet gets "parked", it has more time to reassert its infleunce on the surrounding disc, which leads to it openning the gap there. Part of the material then gets pushed inside the inner boundary. This shows that the final position of the planet could be inside $R_{\rm in}$ for these experiments, if we had a smaller inner boundary. On the other hand, we note that the exact disc structure in the inner disc depends on disc viscosity and other physics, hence the final position of the planets may vary accordingly. 
For the $M_{\rm p} = 12\,\mj$ case, the secondary undergoes rapid inward migration and opens a deep, wide gap extending from $\sim 4$ to $\sim 30$~au. The surface density within the gap is strongly depleted, effectively separating the inner and outer disc. The secondary subsequently stalls at $\sim 15$~au.

\subsection{A grid of planet-secondary injection experiments}\label{sec:Grid}

\begin{figure*}
    \centering
    \includegraphics[width=0.85\textwidth]{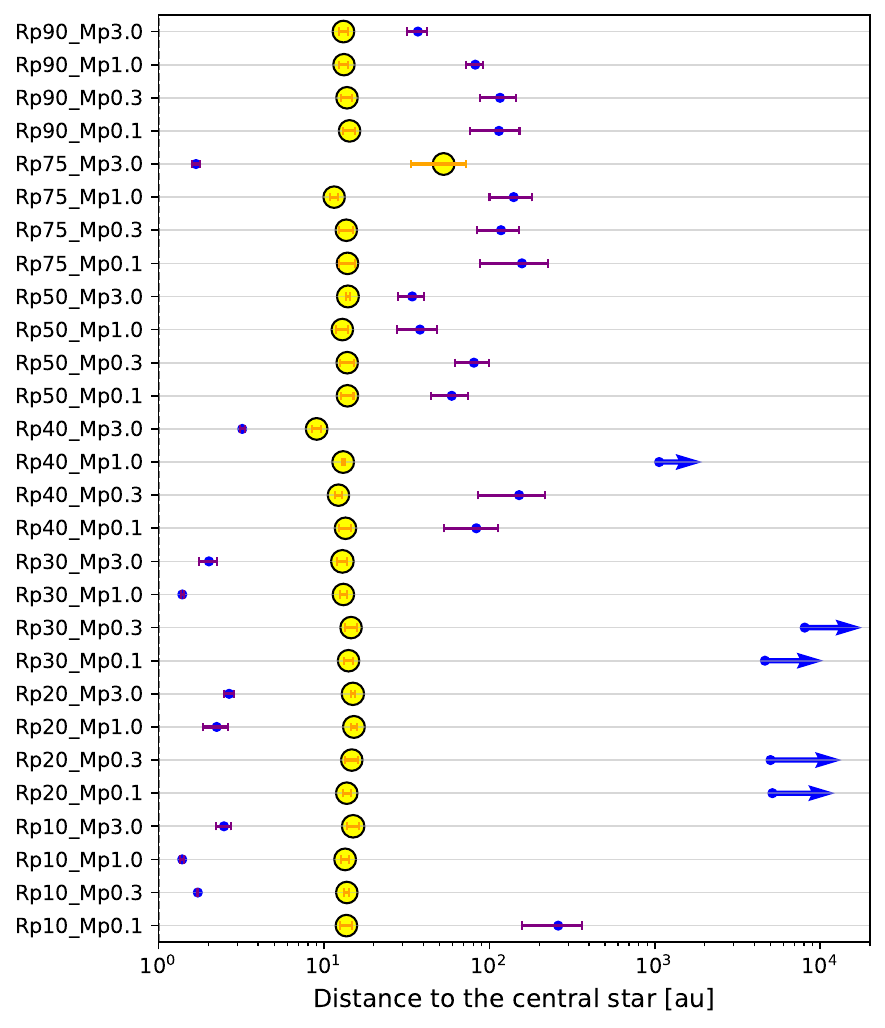}
    \caption{Census of simulated planet-hosting binaries. The y-axis lists simulations labeled as `Rp(initial planet radius in au)\_Mp(initial planet mass in $\mj$)'. Blue dots represent the P1 with purple bars indicating eccentricity, while yellow dots represent the secondary companion with orange eccentricity bars. The size of the yellow dots is proportional to the final secondary mass. Outward directed arrows represent FFPs, with their length proportional to planet velocity.}
    \label{fig:Set1_binaries}
\end{figure*}

We now select the $M_{\rm s} = 12\mj$ case as the most promising one in terms of a tight binary forming in this specific disc. Larger values of $M_{\rm s}$ result in wider binaries as the secondary opens the gap far sooner. On the other hand, smaller values of $M_{\rm s}$ may form even tighter binaries, such as the red curve in the left panel of Fig. \ref{fig:sec-motion}, but these then become only a few times wider than our $R_{\rm in} = 1$~au inner boundary. We would need to decrease the value of $R_{\rm in}$ to follow S-type planet formation in such small binaries.

In this section, we start with the same initial disc as described in \S \ref{sec:single_object}, and inject into it an oligarch with $M_{\rm s} = 12\mj$ and $R_{\rm s}=60$~au at $t=1.4\times 10^4$ years. Simultaneously, we inject one planetary mass object with mass $M_{\rm p} = 0.1, 0.3, 1, 3\mj$ and a range of initial starting radii, $R_{\rm p} = 10,20,30,40,50,75,90$~au at an initial azimuthal angle of $\pi$/2, measured counterclockwise from the secondary, which is initialised at $\phi_{\rm s}=0$. These planetary mass objects affect the disc far less than the secondary does, and hence we may expect that the end result will be a tight binary system with parameters similar to the $M_{\rm s} = 12\mj$ case from \S \ref{sec:single_object}. Here we aim to determine how the planet final position depends on its initial parameters, $M_{\rm p}$ and $R_{\rm p}$.

We run the simulations for additional $ t = 2 \times 10^4 $ years after the objects are injected into the disc. Fig.~\ref{fig:Set1_binaries} displays the resulting orbital configurations for all 28 experiments, labeled as ${\rm RpX\_MpY}$, where X and Y denote the initial radius and mass of the planets, respectively. We find that the secondary evolves similarly across all cases, reaching a final radius of $\sim 15\,$au and a mass $\sim 100\,\mj$. Analysing the figure, we see that planets injected outside the oligarch starting radius, and those just inside it, tend to become wide orbit circumbinary (P-type) planets. In particular, for the 8 planets that started outside of the oligarch, only one was able to migrate past it and become an S-type object.
This can be understood as following. While the oligarch is at the distance of $R\gtrsim 30$~au, its mass is $\lesssim 25\mj$, only a factor $\sim 2$ larger than its initial mass. As is clear from \S 4.2 and Fig. 10 from paper II, only objects more massive than $\sim 0.05 M_*$ are assuredly efficient in launching the planets out of the system. Less massive secondaries increase semi-major axes of the planetary orbits and make them eccentric, but most often are unable to eject them as FFPs. In our simulation here, the oligarch is {\em initially} insuffiently massive to eject the planets. At the same time, by the time it gains sufficient mass to eject the outer planets, they are too far away from it to eject them \citep[cf.][for stability regions]{Holman_Wiegert_99}.

On the other hand, the innermost planets are more likely to survive as S-type planets circling the primary. For $R_{\rm p} =10$~au, only the least massive planet, $M_{\rm p}=0.1\mj$, ends up orbiting outside the binary. This planet migrates inward too slowly, and the secondary catches up with it and scatters outward. All the more massive planets are able to migrate in sufficiently quickly to avoid being destabilised by the secondary; they become S-type planets orbiting the primary.

For intermediate planet injection radii, $R_{\rm p} = 20 - 40$~au, we see that the more massive the planet is, the more likely it is to mature into an S-type planet. Conversely, the less massive the planet is, the more likely it is to be ejected as an FFP. There are exceptions from this rule, of course, due to the stochastic nature of close planet-secondary interactions. Additionally, the less massive the planet is, the larger the eccentricity. This is probably related to lower mass planets being more sensitive to the underlying gravito-turbulence in the disc. Similar qualitative trends can be deduced from \cite{BaruteauEtal11}, their Fig. 4. While their discs were setup differently, and most of their (one per disc) objects eventually migrate to the inner boundary, their lowest mass objects show considerably more stochastic separation evolution than the their most massive ones.

\begin{figure*}
	\centering
	\begin{minipage}{0.33\linewidth}
		\centering
		\includegraphics[width=1\linewidth]{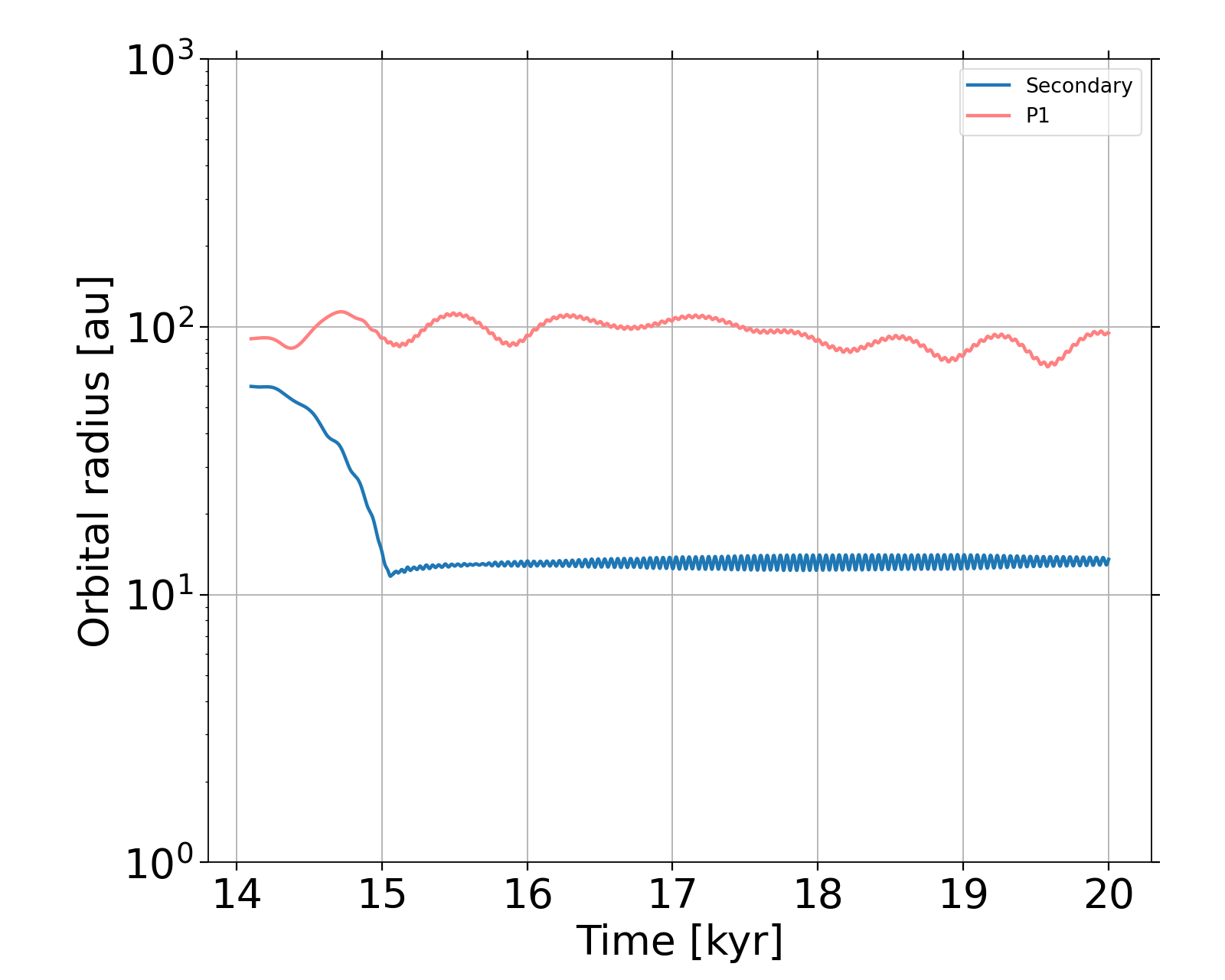}
	\end{minipage}
	\begin{minipage}{0.33\linewidth}
		\centering
		\includegraphics[width=1\linewidth]{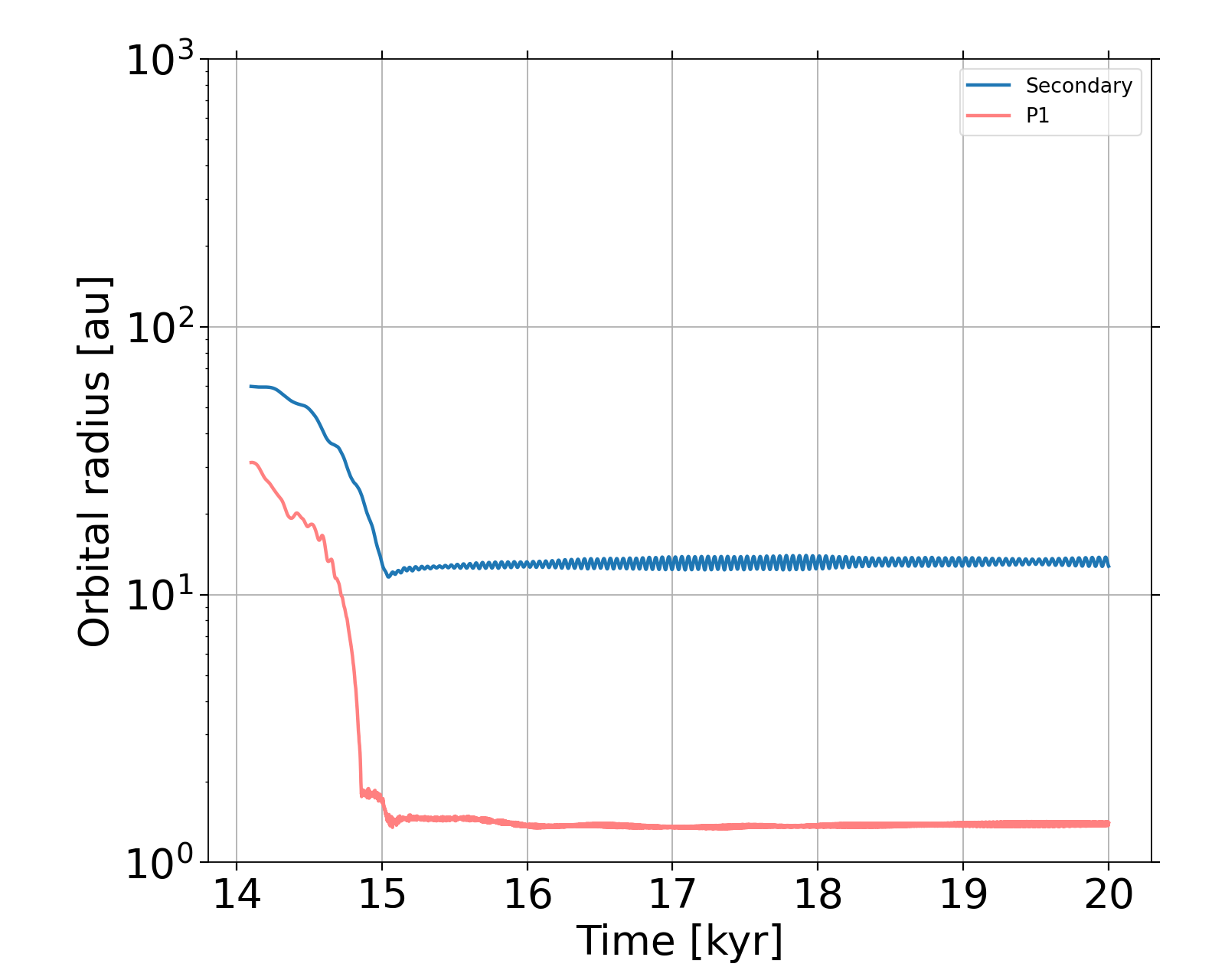}
	\end{minipage}
	\begin{minipage}{0.33\linewidth}
		\centering
		\includegraphics[width=1\linewidth]{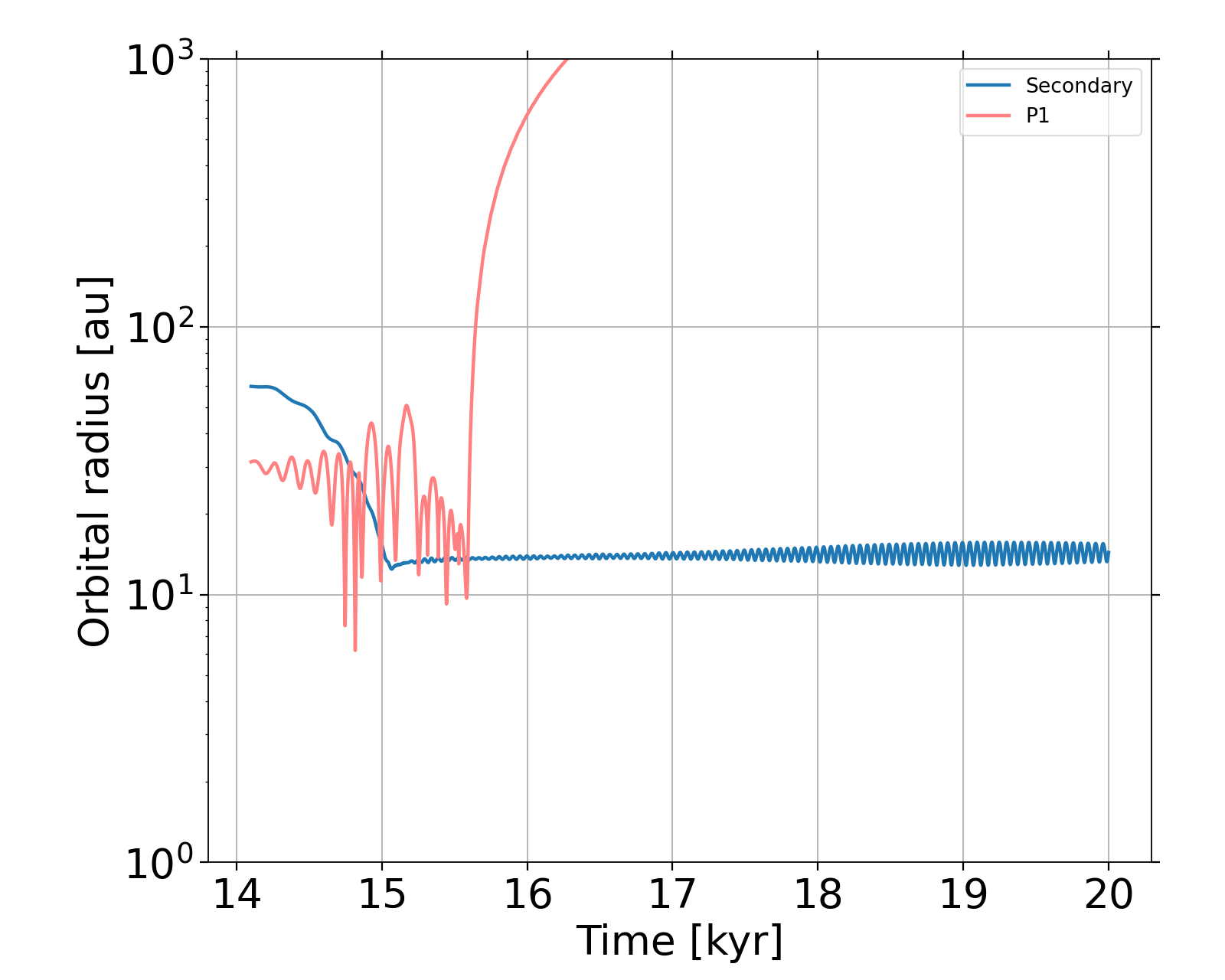}
	\end{minipage}
        \caption{Evolution of the orbital radius for three cases from Fig. \ref{fig:Set1_binaries}: Rp90\_Mp1.0, Rp30\_Mp1.0, Rp30\_Mp0.1. }

    \label{fig:3_cases}
\end{figure*}

We now discuss these results for several specific examples. Fig. \ref{fig:3_cases} illustrates the orbital radius evolution for the planet and the secondary for three representative cases: Rp90\_Mp1.0, Rp30\_Mp1.0, and Rp30\_Mp0.1:

\begin{itemize}
    \item The left panel of Fig. \ref{fig:3_cases}:	Planets starting on the outside of the oligarch are usually unable to match its inward migration speed, therefore a P-type planet configuration results. We find that low-mass planets tend to remain near their injection positions. Higher mass planets may migrate inward eventually (e.g., $M_{\rm p}=3\mj$ migrated in to $R\sim 40$~au, cf. Fig. \ref{fig:Set1_binaries}). However, it is also likely that this migration will be halted by the interactions with the secondary. This could also occur through opening of a wide gap by the secondary, so the P-type planet then sits at the outer edge of the gap in the circum-binary disc. The planet could also be eventually ejected as an FFP if it comes too close to the secondary \citep[in a scenario similar to ][but on much larger spatial scales]{PierensNelson08}, Alternatively, the interaction may scatter the planet inward, as found for Rp75\_Mp3.0. A detailed analysis of the circumbinary and inward-scattered planets is presented in Appendix~\ref{apd:torque}. 

    \item 	The middle panel of Fig. \ref{fig:3_cases}: Here an inner $M_{\rm p}=1\mj$ planet (Rp30\_Mp1.0) manages to migrate inward rapidly enough to avoid close interactions with the secondary, therefore surviving as an S-type planet. Note that this case is reminiscent of the fragmented disc simulation presented in \S \ref{sec:fragmenting_disc}. 

    \item The right panel of Fig. \ref{fig:3_cases}: Here we have a much less massive planet starting at the same initial radius (Rp30\_Mp0.1) as the planet in the middle panel. Since this planet migrates inward much less rapidly than the $1\mj$ one, it is unable to escape a close interaction with the secondary, and is ejected after a few close encounters, similar to the FFP cases studied in paper I and II.
\end{itemize}

\section{Discussion}\label{sec:discussion}

\subsection{Main results}\label{sec:results_discussion}

In this paper we continued to explore the scenario for planet and binary formation by disc fragmentation that we proposed in papers I and II. The key feature of the model is formation of planets by disc fragmentation before the secondary star is born. We envisage that the circumstellar disc around the primary star is in a fragmenting state for $\sim 0.1$~Myr (see \S \ref{sec:the_model}), hatching fragments that tend to migrate inward rapidly. While the fragments are in the planetary regime, the central star binarity status is unchallenged. However, eventually, a massive "oligarch" fragment forms and accretes mass very rapidly to become a low mass star. 
In papers I \& II we found that many of the planets are ejected by close interactions with the rapidly evolving secondary. Here we argued that some of the planets may survive the process and remain bound to either one of the binary components (far more likely the primary), or to the binary as a whole. 

To explore this hypothesis, we performed two types of calculations. In \S \ref{sec:fragmenting_disc}, we presented a proof-of-concept simulation in which the disc was fed by an external mass deposition so that it fragmented. We singled out two particularly massive clumps, $M \sim 6.7$ and $\sim 45\mj$, as the likely progenitors of a gas giant planet and the secondary seed, respectively. This types of modelling differs from that in papers I \& II, where the discs were gravito-turbulent but not fragmenting, and the objects were injected into the disc in physically reasonable orbits, rather than formed ab-initio. We also explored in \S \ref{sec:fragmenting_disc} the dynamics of a $10 \mearth$ solid core injected into the densest regions of the spiral density arms in the disc \citep[motivated by the results of][]{GibbonsEtal14,Longarini_23_solids_collapse_sims}.  The final outcome of the simulation is a tight binary system with separation of $\sim 14\,\mathrm{au}$, eccentricity $e \sim 0.3$, companion mass $\sim 80\,\mathrm{M_J}$, an S-type planet at $a \sim 0.3\,\mathrm{au}$, and an FFP.

We emphasize that, due to numerical limitations, our S-type planets are often in orbits with semi-major axes comparable to the inner boundary of our computational domain. Modelling of the disc on smaller spatial scales and for a much longer duration, possibly taking into accound disc dispersion, is necessary to ascertain where exactly the planets would end up. We expect that many of them would migrate yet closer to the primary, but this depends sensitively on the inner disc evolution in the binary.

In \S \ref{sec:secondary_migration}, we make a first attempt to explore the parameter space of the scenario. We follow an approach similar to that of paper II, injecting collapsed objects in a gravito-turbulent but not fragmenting disc. We experimented with different oligarch starting positions and masses, selecting one case that evolved in a tight binary ($a_{\rm bin}\approx 15$~au and secondary's mass $\sim 0.12 \msun$ (other initial conditions resulted in wider binaries or less massive objects, the cases that we do not focus on here). We then used the same initial disc and the initial oligarch starting radius and mass for simulations in which we additionally injected one giant planet per simulation. We ran a grid of models with planet masses varied from $0.1\mj$ to $3\mj$ and planet's starting location in the range from 10~au to 90~au.

We found that the branching ratio between the survival as an S-type planet and ejection as an FFP is strongly planet-mass dependent. It is the most massive planets that tend to survive as S-types. This has a simple explanation. Planet migration in GI discs occurs via type I migration, so that it is the most massive planets that migrate inward the fastest (see Figs. \ref{fig:sec-motion} \& \ref{fig:3_cases}). Therefore, more massive planets often manage to migrate close in to the primary star quickly. When the oligarch becomes the secondary, it opens a deep gap in the disc, and usually cannot migrate to small radii. This creates a sufficient orbital separations between the secondary and the massive planets that migrated close to the primary. These planets therefore have a decent chance of surviving as S-type planets in binaries. In contrast, low mass planets migrate much more slowly towards the primary. The oligarch has a much greater chance of catching up with them and ejecting them or leaving them as P-type planets on very wide orbits (cf. \S \ref{sec:Grid}).

\subsection{S-type planets in tight binaries}\label{sec:Stype_discussion}

We have shown that disc fragmentation, under the assumptions spelled out in \S \ref{sec:the_model}, can result in formation of S-type planets in tight binary systems. This scenario may apply to systems such as KOI-1257, a binary with $a_{\rm bin} \sim 5.3$~au and eccentricity $e\sim 0.3$, hosting a giant planet with mass of 1.45~$M_{\rm J}$ orbiting the primary at separation of 0.38~au \citep{Santerne14}. Such architecture is challenging for the Core Accretion scenario due to very strong dynamic perturbations of planetesimal orbits by the binary, small mass of the protoplanetary disc, and its short depletion time \citep[e.g.,][]{Venturini_26_planets_in_binaries}. Our scenario avoids such challenges by forming planets very rapidly, through fragmentation of the massive circumprimary disc before the binary forms.

On the other hand, since our model includes binary formation as an ingredient, it must also produce a reasonable field binary population in terms of binarity frequency, binary separation, mass ratio, etc. -- all the while producing the planets as well. Whether it can do so is unclear at the present. We aim to explore these issues in future work.

\subsection{On the mass function of S-type planets}\label{sec:Stype_mass_discussion}

As discussed in the Introduction, the multiplicity rate of hosts of small and large planets/BDs is markedly different. Observations appear to show that while formation of planets less massive than $0.1 \mj$ is strongly suppressed in tight binaries, formation of massive $M_{\rm p}\gtrsim$~a few$\mj$ planets and BDs appears much less suppressed, and may be even enhanced somewhat \citep{Fontanive_19_giants_in_wide_binaries}. This result is counter-intuitive in the context of the CA theory, since gas giant planet formation is more challenging than that of small planets. Gas giant or BD formation requires more mass in both dust (planetesimals) and gas, and takes longer \citep[e.g.,][]{IdaLin04a,MordasiniEtal12}. For example, the detailed population synthesis studies of planet formation in binaries by \cite{Venturini_26_planets_in_binaries,Nigioni_26_planets_in_binaries} find the strongest suppression of planet formation for the most massive planets in tight binaries.

Formation of planets in binaries via GI may naturally explain why the smallest planets are suppressed the most. Note that GI can produce smaller planets, including as low mass as $M_{\rm p} < 1\mearth$, via two different channels, as explained in the Introduction. However, in \S \ref{sec:Grid} we found that even if GI does hatch such planets, those with mass $M_{\rm p} = 0.1\mj$ are unlikely to survive as S-type planets in binaries. This is because planets with mass $M_{\rm p} \ll 1\mj$ migrate in much slower than the oligarch (cf. Fig. \ref{fig:sec-motion}). As overviewed at the end of \S \ref{sec:results_discussion},  all the "low mass" (in the context of GI, $M_{\rm p} \lesssim 0.1\mj$) planets are susceptible to either being ejected by the oligarch (e.g., the left panel of Fig. \ref{fig:3_cases}) or to getting stuck in wide P-type orbital configurations (e.g., the case Rp40\_Mp0.1 in Fig. \ref{fig:Set1_binaries}). 

In contrast, more massive planets, i.e., $M_{\rm p} \ge 1\mj$, stand a much better chance of reaching the safety because they migrate in much more rapidly (e.g., Fig. \ref{fig:sec-motion}). The oligarch is far less likely to catch up with them, as long as they migrate inwards of the \cite{Holman_Wiegert_99} unstable orbits region. Formation of massive S-type planets may therefore be suppressed much less than that of small planets.

Interestingly, Hot Jupiters (HJs), planets with $M_{\rm p}\sim 1\mj$, are intermediate between the rapid and slow planet migration regimes in our simulations. \cite{Ngo_16_Friends_of_HJ} find that these planets have a complex relationship with binaries. The binary fraction of HJ hosts is $\sim 4$ times lower than it is for field stars at separations $a_{\rm bin} = (1-50)$~au. However, the binary fraction of HJ hosts with $a_{\rm bin} > 50$~au is $\sim 3$ times higher than it is for field stars. In other words, very close binaries, $a_{\rm bin} < 50$~au, appear to suppress Hot Jupiters, but wider binaries seem to promote their formation. In the context of our model, this can be understood as following. As any GI planet, HJs would form by disc fragmentation at tens to $\sim 100$~au, and then migrate all the way towards their final positions at $a \lesssim 0.1$~au. In tight binaries, the oligarch outruns the planet and ejects it (e.g., see Rp40\_Mp1.0 in Fig. \ref{fig:Set1_binaries}). The suppression is not total, as planets that already migrated close enough to the primary survive as S-type ones (e.g., Rp20\_Mp1.0 in Fig. \ref{fig:Set1_binaries}), but the probability of this is small because such planets migrate slower than super-Jupiters. On the other hand, in very wide binaries, the oligarch has not migrated inwards much. Here, the triggered fragmentation scenario \citep{Cadman_22_triggered_fragm_binary} may account for more HJs in wide binaries. 

There is also an interesting analogy to the statistics of very close stellar binary systems \citep{Tokovinin_06_tripples}: 96\% of binaries with periods of less than 3 days have an additional stellar companion at wider separations. As pointed out by \cite{FC_BG_21}, a very substantial disc mass is required to power inward migration of the secondary in such systems. Such discs may also have enough mass to form the tertiary companions, explaining why the closest binaries are often in triple systems. It is possible that the same scenario applies for massive planets and BDs in binary systems.

\subsection{Comparison to the FFP mass function}\label{sec:FFP_discussion}

Our model makes an explicit link between FFPs and planets within the tight binary systems. The former are those that were ejected from the binary and the latter are those survived on bound orbits aroud one of the stars. Due to the strong mass dependency of the ejection process, our scenario predicts that the mass function of the FFPs must be more bottom heavy than the one for planets in binaries. The observations probably support this prediction.

The mass function of planets in tight binaries is not yet well known due to strong selection bias for such systems \citep{Moe_21_planets_in_binaries,Thebault_25_planets_in_binaries}. However, observations show that small planets are suppressed more in tight binaries than gas giant planets (cf. the previous section).  Thus it is likely that the mass function of planets in binaries is less steep than that of planets in single stars. \cite{Zang_25_bound_microlensing} shows that the microlensing population of bound planets in single stars can be represented by two Gaussians peaked at masses of $\sim 7.4\mearth$ and $\sim 770\mearth$, and that there is a total of $\sim 0.57$ and $\sim 0.053$ planets per Mdwarf star in these peaks. Following our logic, the ratio of super-Earths abundance to that of super-Jupiters in tight binaries should not exceed $\sim 10$, as in the \cite{Zang_25_bound_microlensing} mass function.  

The FFP mass function approximately scales as $dN/dM_{\rm p} \propto 1/M_{\rm p}$ \citep[see the review by][]{Mroz-23-Microlensing-planets-review}. For such a mass function, the ratio of numbers of planets with the above masses would be $\sim$ the inverse of the mass ratio, that is $\sim 100$. So, in this admittedly preliminary analysis, there appears to be an order of magnitude more super-Earths per gas giant in the FFPs than in the bound planet populations.

\subsection{Circumbinary planets in tight binaries}\label{sec:Circumbinary_discussion}

Although not an intended outcome of this study, our model produces a substantial population of planets on circumbinary orbits. These planets are generally expected to be of low mass, as lower-mass planets migrate more slowly and can be more readily halted by the secondary. However, as shown in \S\ref{sec:Grid}, even planets with masses exceeding $1\mj$, which would normally be expected to migrate inward rapidly, can remain on wide orbits following the injection of the secondary.

This behaviour is reminiscent of the results reported by Paper I \& II, where we found that planets embedded in gravito-turbulent discs can be scattered onto wide orbits through interactions with a secondary object and subsequently remain there for extended periods. Additionally, interactions with dense self-gravitating structures in the disc can also generate outward migration for lower mass objects \citep[see][]{BaruteauEtal11,Rowther_Meru_2020}. 

In the simulations presented here, close encounters between the planet and the secondary usually fail to eject the planet when it is initialised on an orbit outside of the secondary. This is different from the results of papers I \& II, where planets were initialised interior to the secondary and were more frequently ejected. Additionally, in the simulations shown in Fig.~\ref{fig:Set1_binaries}, close interactions between planets and the secondary typically occur before the latter has accreted a substantial amount of mass. Due to this, the secondary's kick to the planet in such interactions is weaker. Instead of a clean ejection, the planet is usually scattered onto a wider orbit, making further close interactions less likely. In many cases, the planet eventually becomes trapped near the outer boundary of the unstable region, allowing it to survive on a wide circumbinary orbit for a prolonged period. A detailed analysis of the migration histories and disc torques experienced by these circumbinary planets is presented in Appendix~\ref{apd:torque}. 

% \textbf{Observationally, Fig.~6 in \cite{Thebault_25_planets_in_binaries} shows that many circumbinary planets lie close to the instability boundary, qualitatively consistent with the characteristics of our simulations. However, most of these observed circumbinary planets are found in systems with binary separations $\le 1$~au. }

In this research, we have presented a preliminary statistical analysis of the surviving circumbinary planets, but a more detailed investigation is required. On the one hand, the limited duration of our simulations may not capture the long-term dynamical evolution of these systems. On the other hand, the finite resolution and restricted spatial extent of the disc simulations may also influence the results. Consequently, some of the circumbinary planets identified here may eventually be ejected as FFPs on longer timescales, or alternatively migrate inward and become S-type planets. Future studies with longer simulation times and a broader exploration of parameter space will be necessary to establish the fate and occurrence rate of circumbinary planets more robustly.

\subsection{Suppression of planet formation in wide binaries}\label{sec:wide_binaries_discussion}

The distribution of binary separation for planetary hosts in \cite{Thebault_25_planets_in_binaries} shows that planet formation remains somewhat suppresed out to surprisingly wide separations, $a_{\rm bin}\sim 500$~au. At the first glance, neither Core Accretion nor the GI scenario proposed here can explain this result naturally. \cite{Venturini_26_planets_in_binaries}, cf. their Fig. 6, find that binaries wider than $\sim 100$~au do not suppress planet formation. This is expected. In the CA picture, planets form effectively at distances of a few to $\sim 10-30$~au \citep[e.g., Fig. 7 in][]{Burn_21_BernCA_lowMstar}.  If circumprimary disc is cutoff at the distance of $\sim (1/4) a_{\rm bin}$, then planet formation is mainly unaffected for binaries wider than 100~au \citep[except for gas giants, see][]{Venturini_26_planets_in_binaries}. 

In the scenario we propose here, an oligarch-turn-secondary at several hundreds au is also not likely to affect the planets forming by GI. Indeed, GI planet birthplace is $\sim 50-100$~au \citep[e.g.,][]{Rafikov05,Clarke09,KratterEtal10}, and they migrate in rapidly. Most of these planets would not be affected by a non-migrating secondary at hundreds of au\footnote{To test this, we set up simulations with disc more extended than in \S \ref{sec:single_object}, and injected oligarchs at separations as large as 270 au. It was found that the binary separation shrinks rapidly to $a_{\rm bin} \lesssim 100 $~au. It appears difficult to produce wide binaries by disc fragmentation.} It is possible that binaries significantly wider than $\sim 100$~au do not form by circumprimary disc fragmentation. These binaries could form by filament or core fragmentation on two {\em initially} quasi-independent stars surrounded by their own discs \citep[e.g., see the Introduction, and][]{Offner_23_binaries_review}. If this were the case, then a 400 au binary would cut the circum-primary disc at $\sim 100$~au, depleting the disc mass at GI-prone radii rapidly. It is quite likely that GI planet formation is then weakened but not completely stifled.

\subsection{Individual notable systems}\label{sec:systems_discussion}

\subsubsection{HD~87646}

HD~87646A is 1.12 $\msun$ primary star hosting a $12.4\,\mathrm{M_J}$ planet at $0.12\,\mathrm{au}$ and a $57\,\mathrm{M_J}$ BD at $1.6\,\mathrm{au}$. Both the BD and the companion star (at $a_{\rm bin}\approx 20$~au) are eccentric, with $e \sim 0.5$ \citep{Fontanive_19_giants_in_wide_binaries,Ma16}. The circumprimary disc in the system would be cutoff at $R\sim 3-4$~au, depending on the disc viscosity \citep{Artymowicz94}. Let us assume that formation of the planet and the BD occurred within a circumprimary disc with mass at least a few times the sum of the planet and the BD mass. We arrive at a minimum $M_{\rm d} \gtrsim 0.2 \msun$ disc with outer edge of just a few au. Such a disc would be gravitationally unstable. GI-driven turbulence \citep[e.g.,][]{Gammie01,LodatoRice05} would result in the disc depletion time that can be estimated as $\lesssim 10^3$~years only. It would be astonishing if the super-Jupiter and the massive BD could form via CA in such conditions. 

Note that GI also cannot form this remarkable system in {\em its current configuration}, i.e., neither the planet nor the BD could have formed via GI in the current locations. Very massive but compact discs with $Q < 1$ will generate GI-driven trubulence and heating, resulting in the discs puffing up and self-regulating to $Q\gtrsim 1$ \citep[][]{Lin87,Gammie01,LodatoRice04}. For the disc to be not only self-gravitating but to also fragment on isolated objects, it must also cool rapidly, which is only possible at wide separations, $R\gtrsim 50$~au, where the disc is less optically thick \citep{Rafikov05,Levin07,Clarke09,LeeH_25_dusty_GI}. 

We believe that HD~87646 system presents us with a solid proof that planets and BDs in S-type configurations can form {\em before} the binary attains its final configuration. As we argued above, forming HD~87646 planet/BD in the present binary is not possible either by CA or GI. Making them at $R > 50$~au by GI is possible, perhaps even in the current system configuration. However, the planet/BD would then have to migrate past the secondary in the system without getting ejected. We do not think this is likely.

%In contrast, forming the planets, and the BD, at $\gtrsim 50$~au via disc fragmentation stretches no physical constraints. We only need both of the objects to migrate into the inner disc by the time the seed of the secondary star was born in this system. 

\subsubsection{HD~72892}

The system is composed of a $\sim 1\msun$ primary and a $77\mj$ secondary on the boundary between BDs and low mass stars. The system separation and eccentricity is 8~au and 0.38, respectively. HD~72892b is $M_{\rm p}\approx 5.5\mj$ planet orbiting the primary at separation of 0.23~au, with eccentricity $e\approx0.4$. Using the fitting equations for disc trucation from \cite{Venturini_26_planets_in_binaries}, we estimate the outer edge of the circumprimary disc would be at $R\approx 2.2$~au. Forming the super-Jupiter via CA in such a small disc is difficult, given the high eccentricity of this system, and a short disc dissipation time that we estimate at $\lesssim 10^5$~years even for viscosity parameter of $\alpha=10^{-3}$.

It is interesting to note that the this system is somewhat similar to the HD~87646's inner two objects, minus the secondary star. It is possible that HD~72892's disc was less massive than the one in HD~87646, and so did not hatch a more massive star, but otherwise the evolution of the super-Jupiter and a very massive BD in these two systems were similar.

\subsubsection{HD~41004}

HD~41004 is a spectacular binary of two stars with masses of $0.7$ and $0.4\msun$, separated by 23 au, in which both the primary and the secondary host S-type companions. The primary is orbited by a 2.5~$M_{\rm J}$ planet, whereas an 18~$M_{\rm J}$ BD orbits the secondary star at a separation of only $0.02$~au \citep{Zucker03,Zucker04,Thebault_25_planets_in_binaries}.  Focusing on just the secondary star here, we find that the outer edge of its protoplanetary disc would be at $R < 6$~au. Forming the BD via CA in such a small yet massive disc would be most unusual, especially given that the secondary star is an M-dwarf. For example, population syntheses of CA planet formation shows very few gas giant planets ($M_{\rm p} > 0.1\mj$), and certainly no BDs, for M-dwarfs even in discs around single stars \citep{Miguel_20_CA_Mdwarfs,Mulders_21_CA_Mdwarfs,Burn_21_BernCA_lowMstar}. 

It is possible that the secondary star in this system was able to capture a BD that formed by fragmentation of the circumprimary disc. Alternatively, the system could have been much larger at birth, i.e., if it formed via a filament fragmentation. The BD could then have formed in an initially larger disc of the secondary star to begin with. Dissipative gas dynamics during binary system growth \citep[e.g.,][]{Tokovinin_Moe_20_binaries} could then have brought the two stars much closer than it was previously.

\section{Acknowledgement}

LZ is supported by China Scholarship Council (CSC) in collaboration with the University of Leicester.  

This research used the ALICE High Performance Computing facility at the University of Leicester, and the DiRAC Data Intensive service at Leicester, operated by the University of Leicester IT Services, which forms part of the STFC DiRAC HPC Facility (\href{www.dirac.ac.uk}{www.dirac.ac.uk}). 

\section{Data availability}

The data obtained in our simulations can be made available on reasonable request to the corresponding author.

%%%%%%%%%%%%%%%%%%%% REFERENCES %%%%%%%%%%%%%%%%%%

\bibliographystyle{mnras}
\bibliography{nayakshin,Luyao}

\appendix                  

\section{Tests with varying masses of planets and secondaries in the fragmenting disc}
\label{apd:test} 

In \S~\ref{sec:fragmenting_disc}, we injected planet P1 with mass $M_{\rm p1} = 1\,\mj$ and the secondary embryo with mass $M_{\rm s} = 5\,\mj$ into the respective gas clumps to study system evolution. While these values for the post-collapse objects inside GI clumps are physically reasonable, it is important to explore the sensitivity of results to the initial masses of P1 and S1.

\begin{figure}
    \centering
    \includegraphics[width=0.45\textwidth]{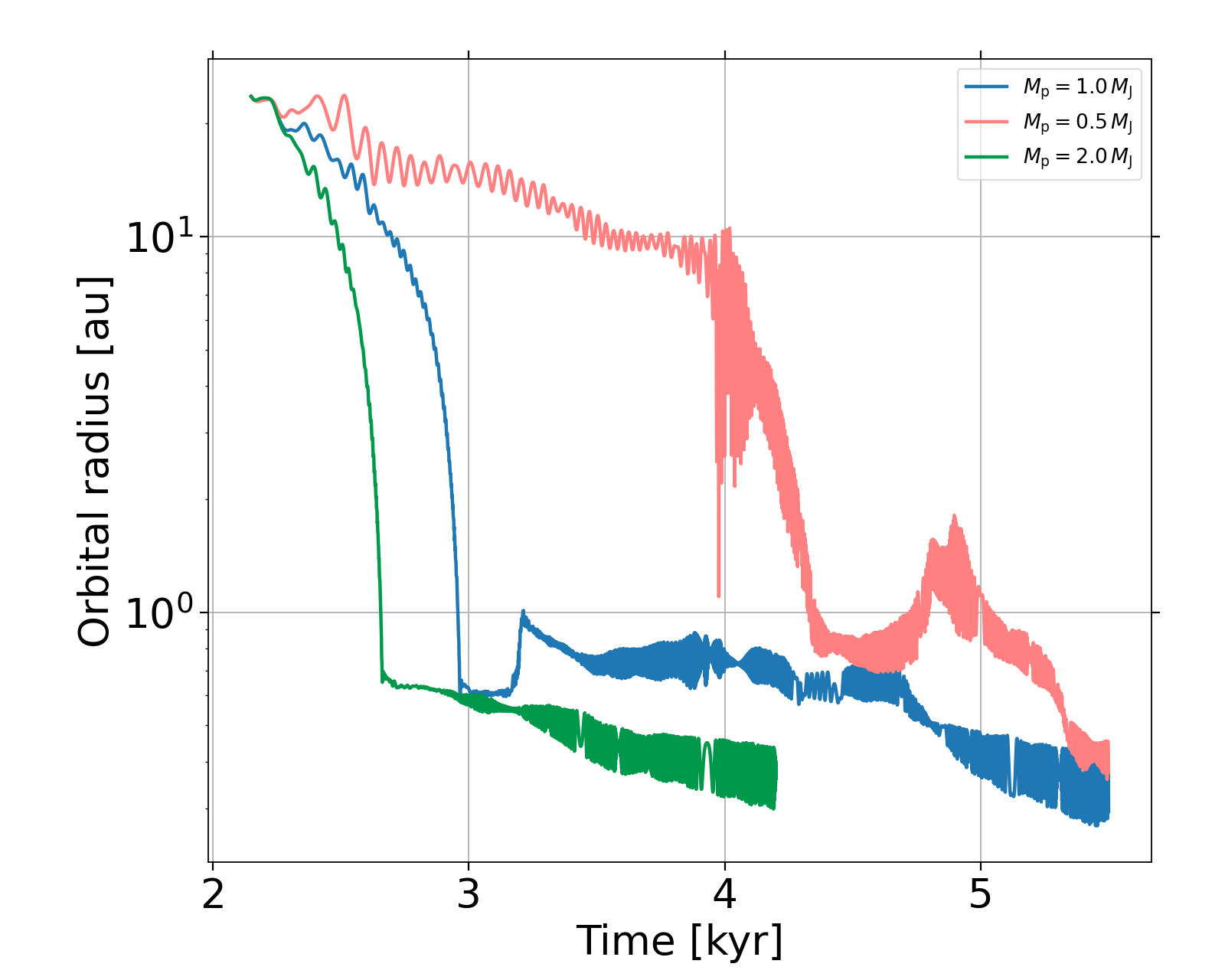}
    \caption{Orbital radius evolution of P1 with masses of $M_{\rm p1} = 0.5, 1, 2\mj$. }
    \label{fig:t-mr-pla}
\end{figure}

We first consider P1's orbital evolution separately. Following the set up described in \S~\ref{sec:fragmenting_disc}, we performed three separate simulations with P1 mass being $M_{\rm p1} = 0.5, 1, 2\mj$. Fig. \ref{fig:t-mr-pla} shows orbital evolution of the planet for these three different cases. As expected for the type I migration regime, Fig. \ref{fig:t-mr-pla} shows that the higher the mass of the object, the faster is its migration. Additionally, the lower the mass of the object, the more stochastic is its migration in the gravito-turbulent disc \citep[e.g., see examples in][]{BaruteauEtal11}. That said, the final outcome of the simulations is similar: the host fragment is tidally shreaded but the planet migrates rather rapidly close to the inner edge of our disc. 

\begin{figure}
    \centering
    \includegraphics[width=0.45\textwidth]{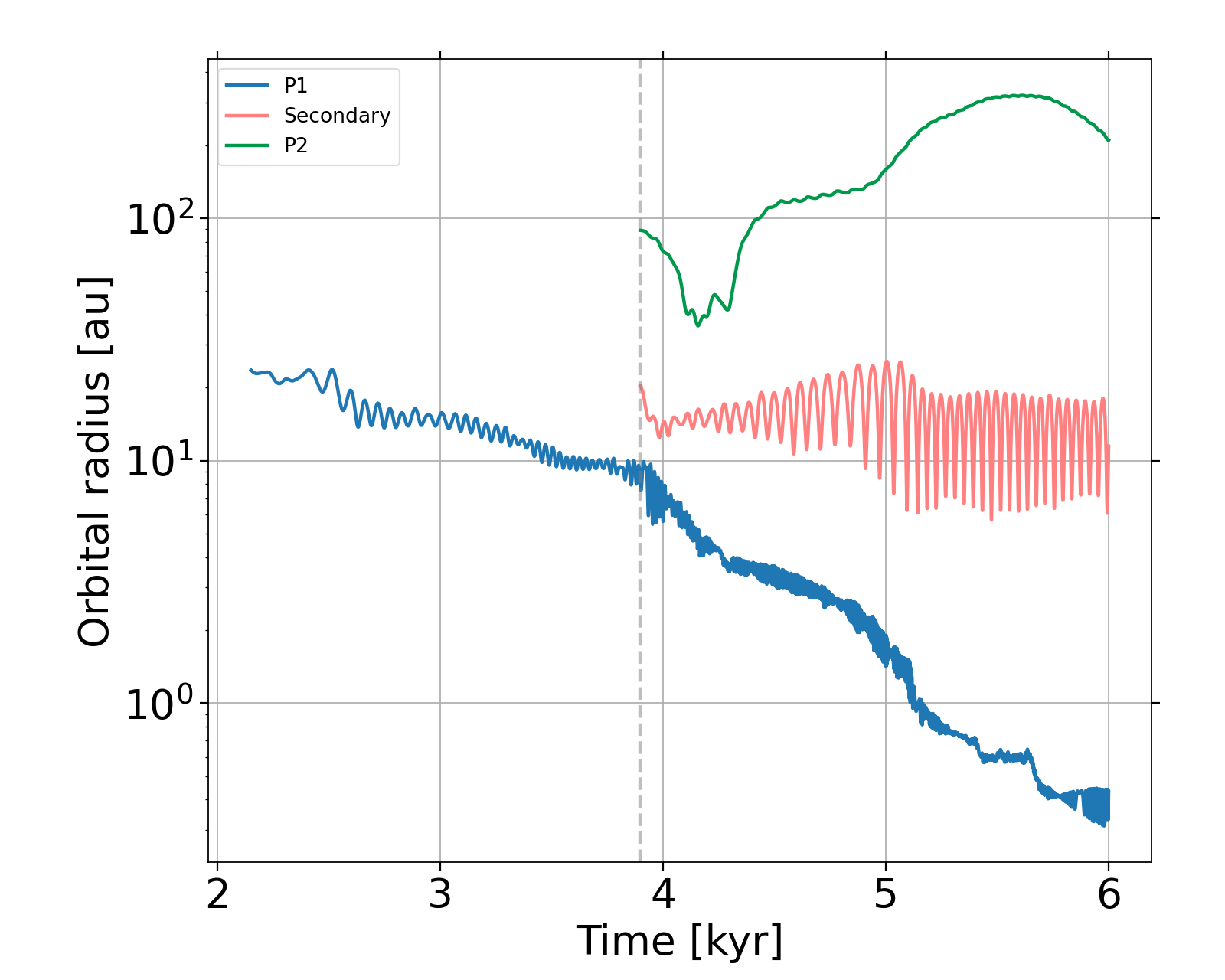}
    \caption{Orbital radii of the secondary and planets in the simulation of $M_{\rm p1} = 0.5\,\mj$. The grey dotted line marks $t=3.9\,\mathrm{kyr}$, when the secondary and the other planet was injected. }
    \label{fig:rt_p0.5}
\end{figure}

In the case of $M_{\rm p1}=0.5\,\mj$, P1 lingers at $\sim 10$~au much longer than in the other two cases. It may therefore be possible that the secondary, injected just outside of that location, would dislodge P1 and eject it to become an FFP. To study this further, we inject the secondary seed ($5\,\mj$) into the oligarch clump, and the low mass P2 into an exterior clump at $t = 3.9\,\mathrm{kyr}$ (repeating the approach of \S~\ref{sec:fragmenting_disc}). Fig.~\ref{fig:rt_p0.5} presents orbital evolution of the three objects, showing that P1 does continue its inward migration. While its dynamical evolution differs somewhat from that in \S~\ref{sec:fragmenting_disc}, P1 ultimately still becomes an S-type planet in a tight binary system. Finally, P2 in this simulation is not ejected into an FFP, becoming a wide orbit P-type planet.

% the higher-mass P1 ($2 \mj$), migration is rapid and stable inward, while lower-mass P1 ($0.5\mj$) migrates slowly and stochastically. The clump embedded P1 is tidally disrupted quickly. After the clump is shredded, the motion of P1 resembles that of objects in Fig. \ref{fig:sec-motion}, where low-mass planets linger in wide orbits. However, low-mass planets are prone to disturbances from external spirals and clumps. After being accelerated by a clump oligarch, it is ejected inward at near 4 kyr. 

\begin{figure}
    \centering
    \includegraphics[width=0.45\textwidth]{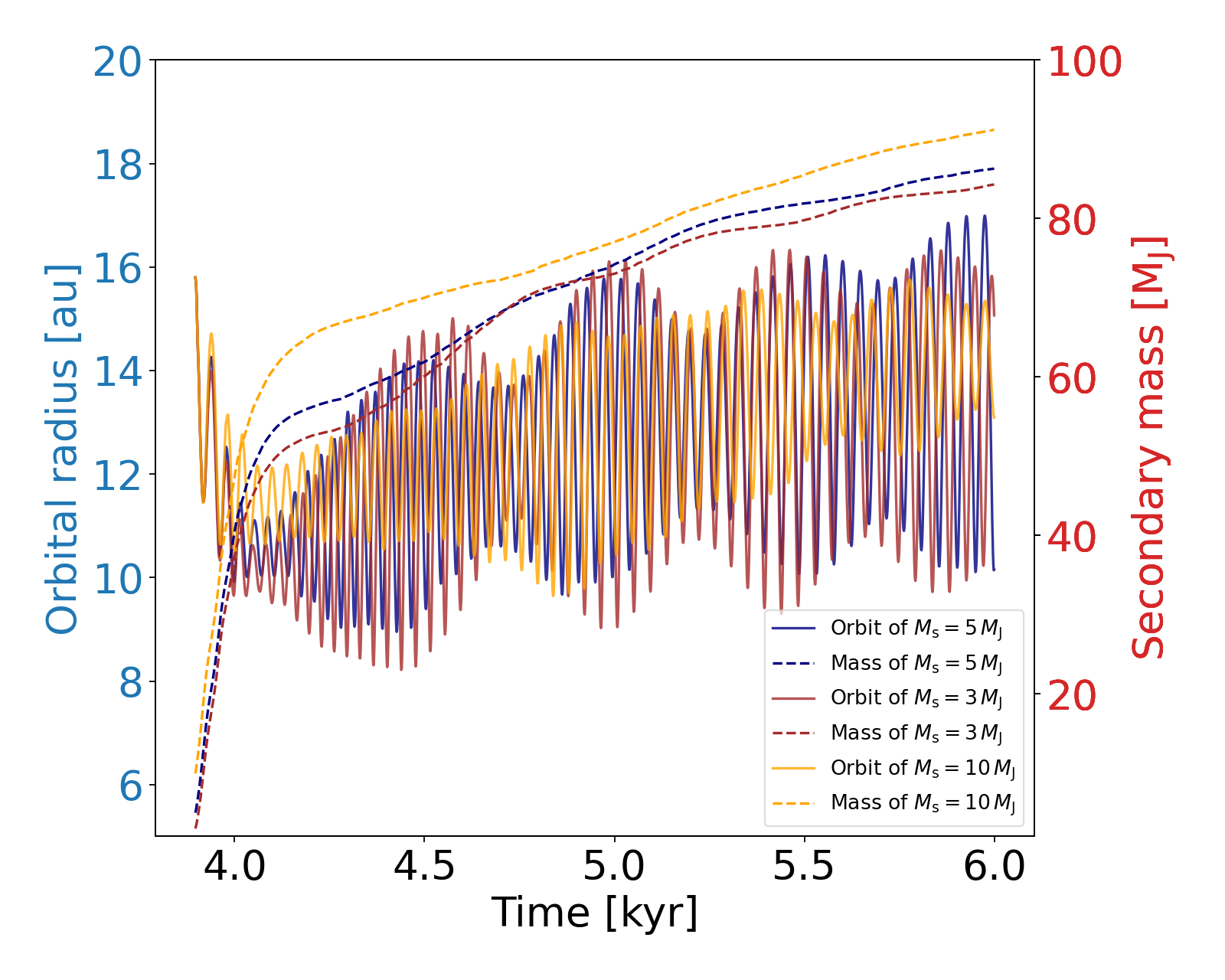}    
    \caption{Time evolution of the orbital radius and mass of the secondaries with masses of $M_{\rm s} = 3, 5, 10\mj$. }
    \label{fig:mr-sec}
\end{figure}

Finally, we additionally explore the injection of secondaries with varying masses $M_{\rm s} = 3, 5, 10\mj$ into the  $M_{\rm p1} = 1\,\mj$ at $t = 3.9\,\mathrm{kyr}$ simulation from \S \ref{sec:fragmenting_disc}. As shown in Fig.~\ref{fig:mr-sec}, these cases exhibit qualitatively similar evolutionary behavior for the secondary star. We do not show the outcomes for P1 and P2 for brevity here. In all three cases, P1 became survived as an S-type planet in the binary system. This shows that our model is relatively insensitive to the values of the masses of P1 and the initial mass of S1.

\section{Analysis of circumbinary planets in Non-Fragmenting discs}
\label{apd:torque} 

As discussed in \S~\ref{sec:Grid}, planets initially injected just outside the secondary tend to remain on wide orbits. This is in contrast to the corresponding single-star simulations with similar initial planet parameters in \S~\ref{sec:single_object}, where, e.g., a $1\,\mj$ planet undergoes rapid inward migration (Fig.~\ref{fig:sec-motion}). However, when the secondary is present, the migration rate of the planet is significantly reduced, and the planet may even stall entirely, as seen in the Rp90\_Mp1.0 case shown in Fig.~\ref{fig:3_cases}. To further understand this, we consider three representative cases in the non-fragmenting disc: a planet that is ultimately as close to the secondary as the \cite{Holman_Wiegert_99} orbital stability analysis allows (Rp50\_Mp1.0); a planet that survives on a wide circumbinary orbit well outside the instability region (Rp75\_Mp1.0); and a planet that is scattered inward of the secondary's orbit (Rp75\_Mp3.0).

\begin{figure*}
	\centering
	\begin{minipage}{0.45\linewidth}
		\centering
		\includegraphics[width=1\linewidth]{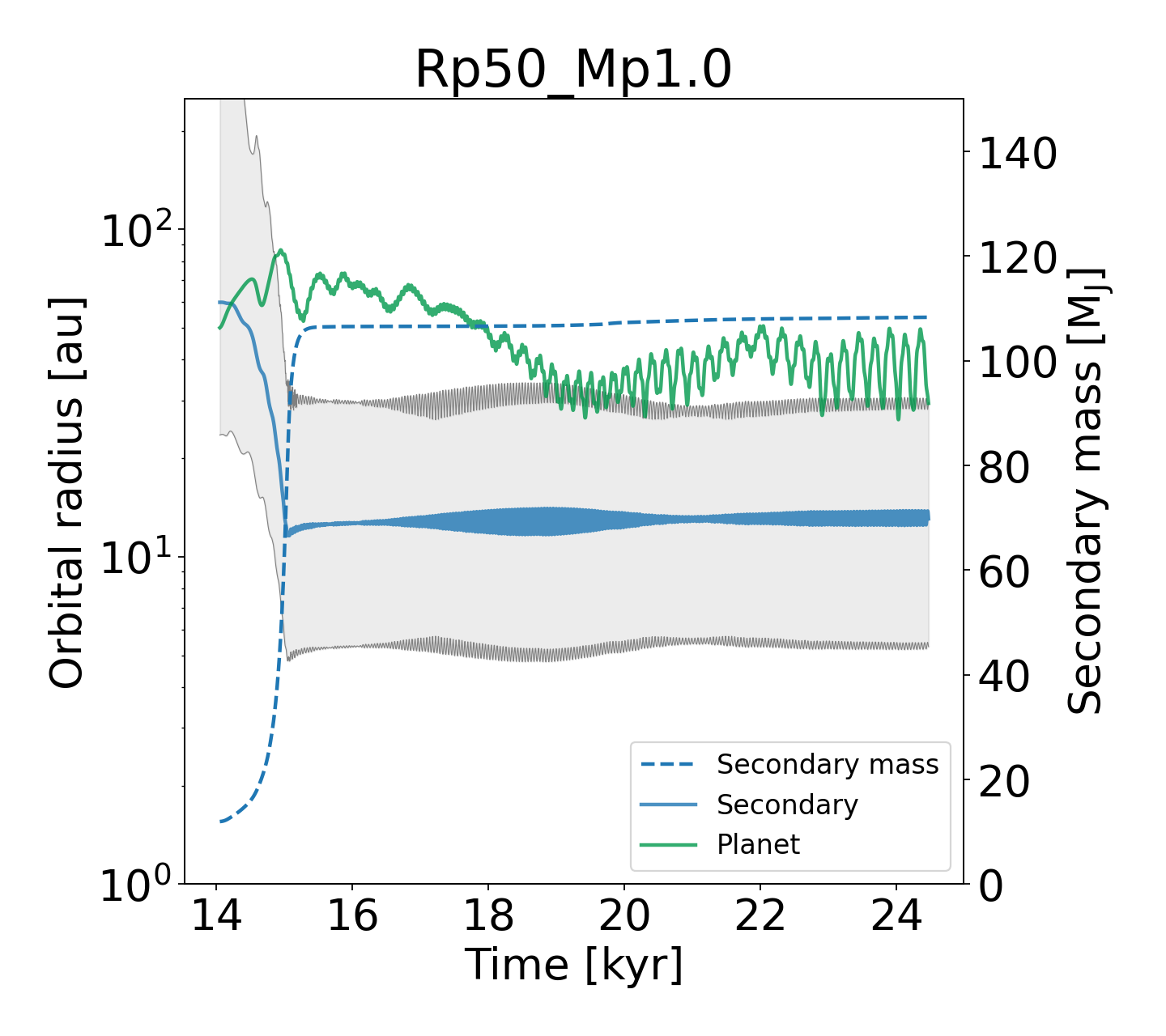}
	\end{minipage}
	\begin{minipage}{0.38\linewidth}
		\centering
		\includegraphics[width=1\linewidth]{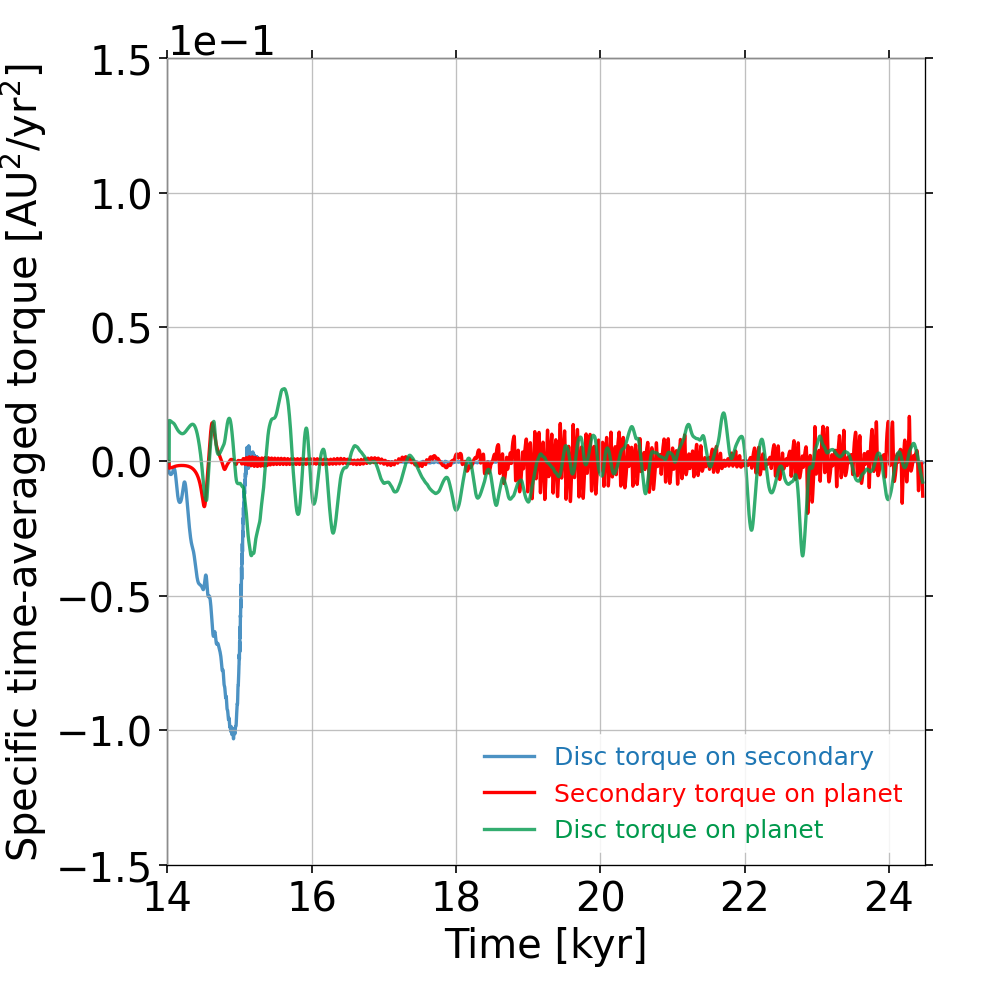}
	\end{minipage}

	\begin{minipage}{0.45\linewidth}
		\centering
		\includegraphics[width=1\linewidth]{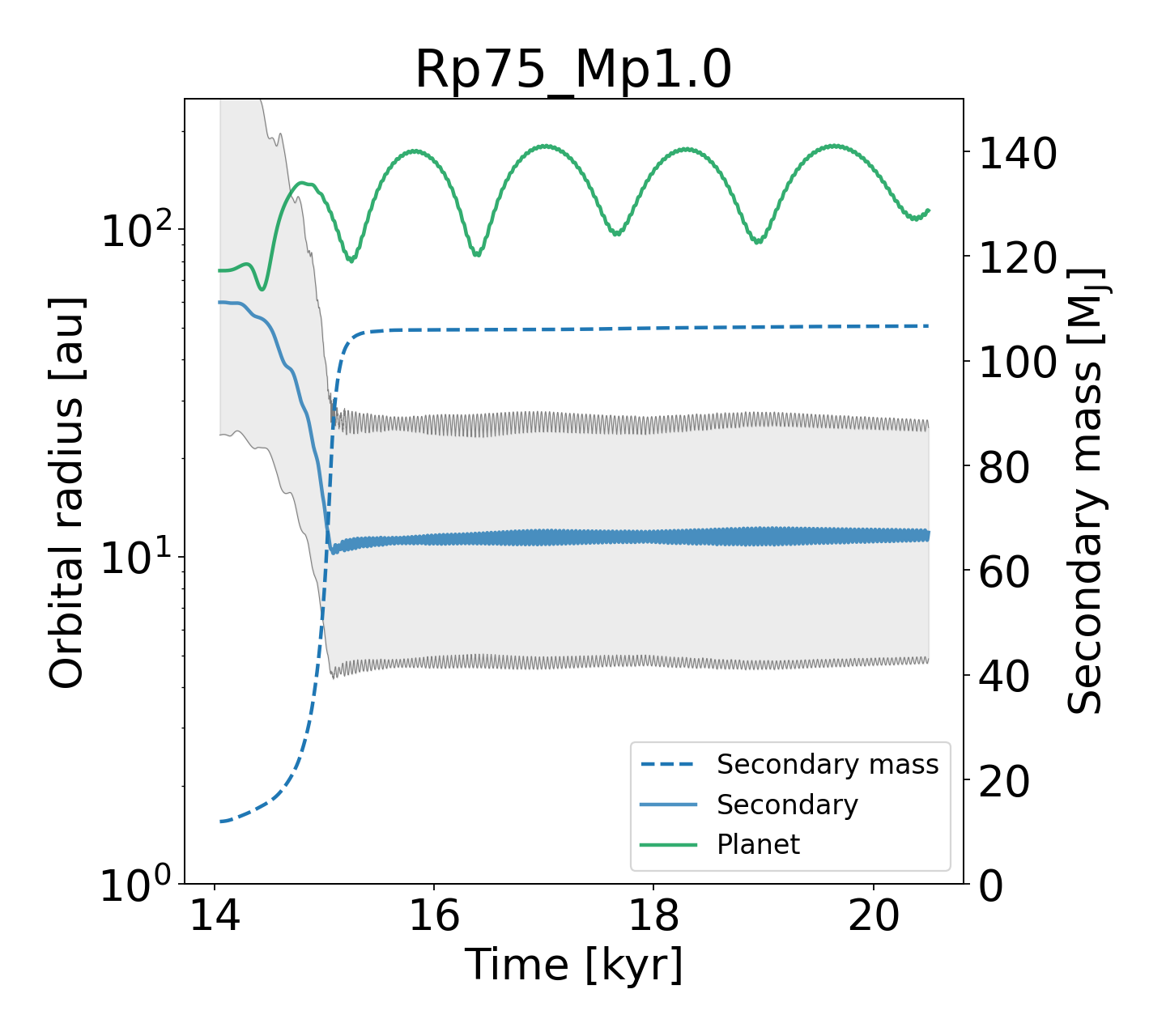}
	\end{minipage}
	\begin{minipage}{0.38\linewidth}
		\centering
		\includegraphics[width=1\linewidth]{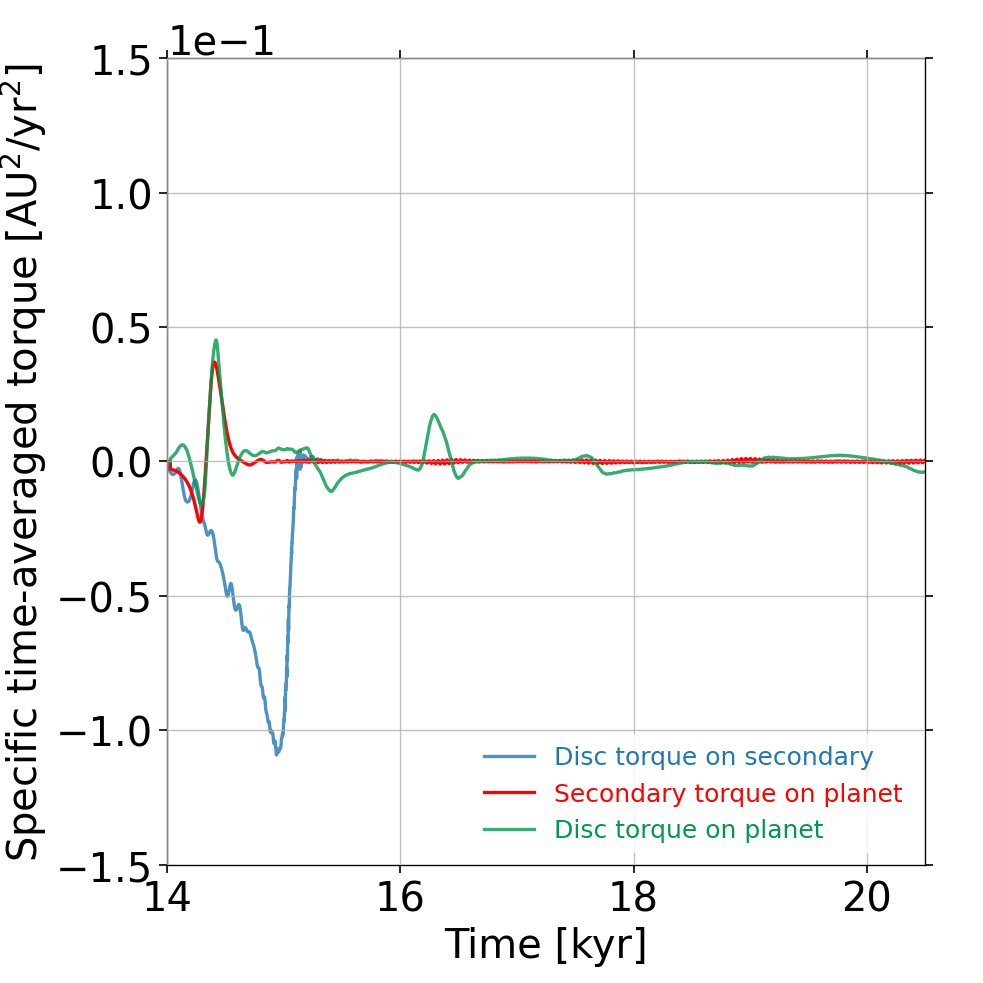}
	\end{minipage}

	\begin{minipage}{0.45\linewidth}
		\centering
		\includegraphics[width=1\linewidth]{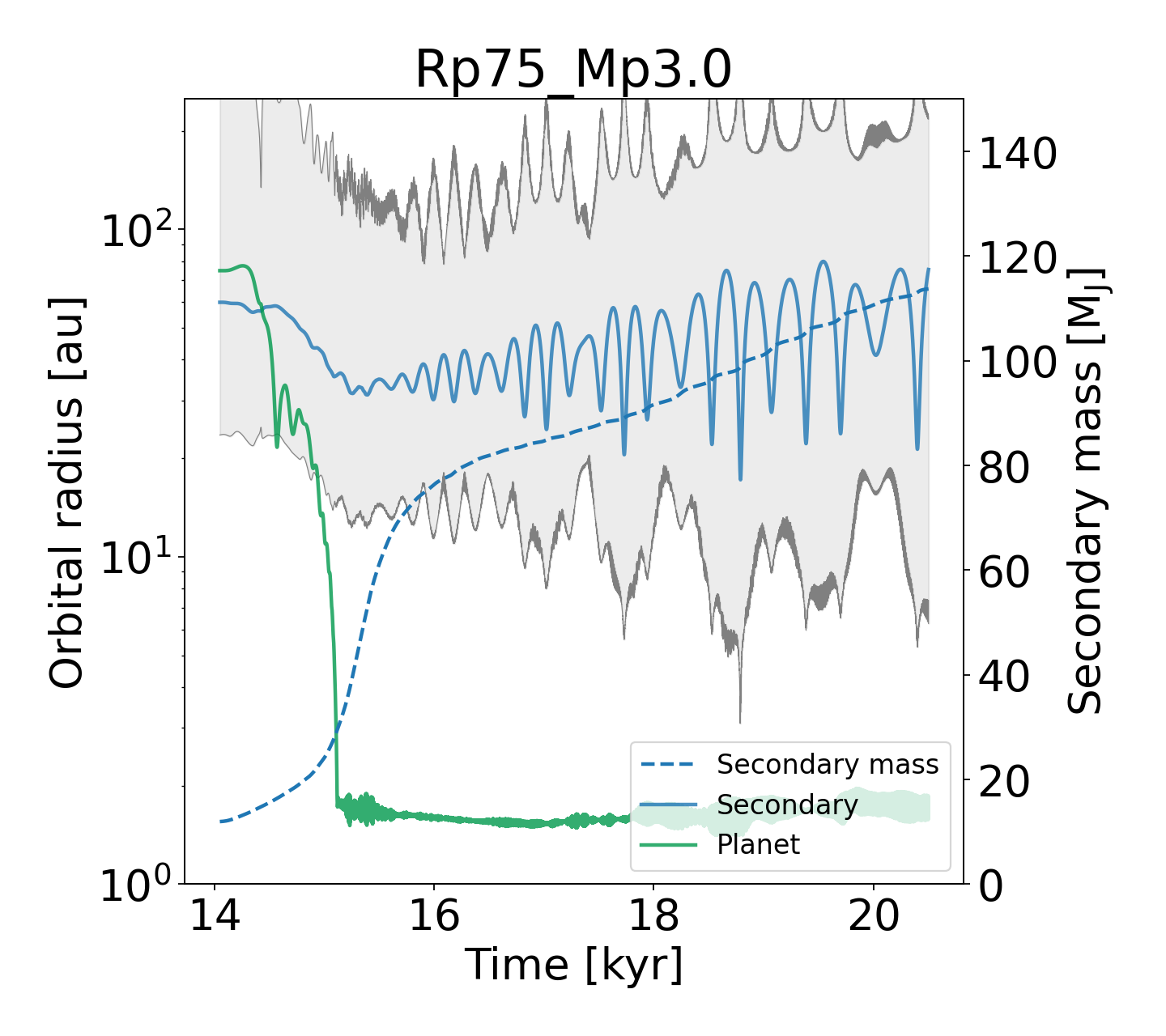}
	\end{minipage}
	\begin{minipage}{0.38\linewidth}
		\centering
		\includegraphics[width=1\linewidth]{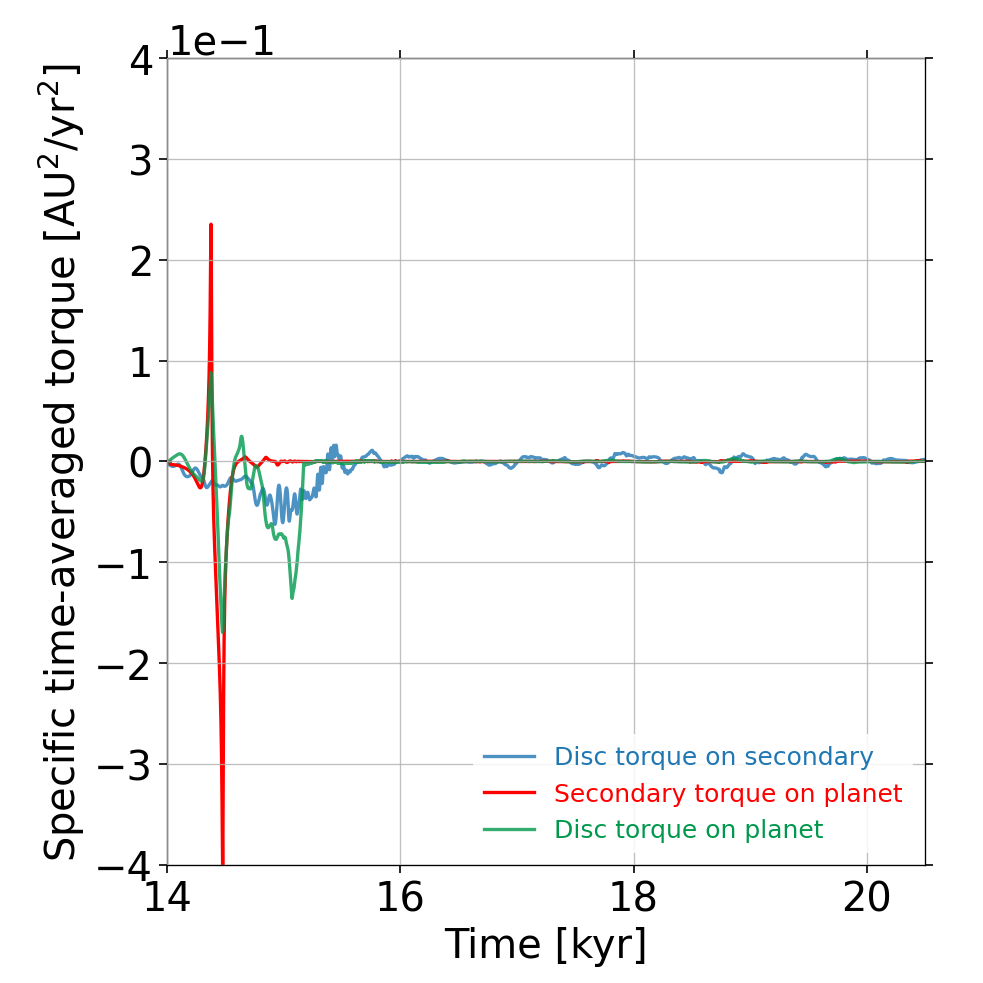}
	\end{minipage}
        \caption{
        Left: Time evolution of the orbital radii of the objects (solid curves) and the mass of the secondary (dashed curve) for the Rp50\_Mp1.0, Rp75\_Mp1.0, and Rp75\_Mp3.0 simulations. The grey shaded region denotes the radially unstable zone predicted by the criterion of \citet{Holman_Wiegert_99}. Right: Time evolution of the specific torques exerted on the objects, averaged over a 100-year time window. For the case of Rp50\_Mp1.0, there is an expansion of time axis to show the details of the planet's orbital evolution; for the case of Rp75\_Mp3.0, the y axis is expanded to show the magnitude of the torques. }

    \label{fig:torque_mr}
\end{figure*}

Fig.~\ref{fig:torque_mr} presents the radial positions of the planet and the secondary, as well as the disc and the secondary's torques on the planet. We also present the separation between the planet and the secondary as a function of time in Fig.~\ref{fig:r0}.

\begin{figure*}
	\centering
	\begin{minipage}{0.33\linewidth}
		\centering
		\includegraphics[width=1\linewidth]{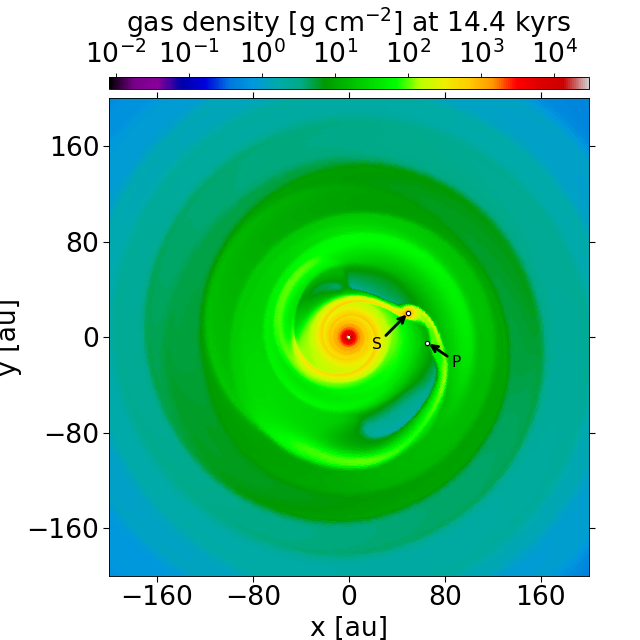}
	\end{minipage}
	\begin{minipage}{0.33\linewidth}
		\centering
		\includegraphics[width=1\linewidth]{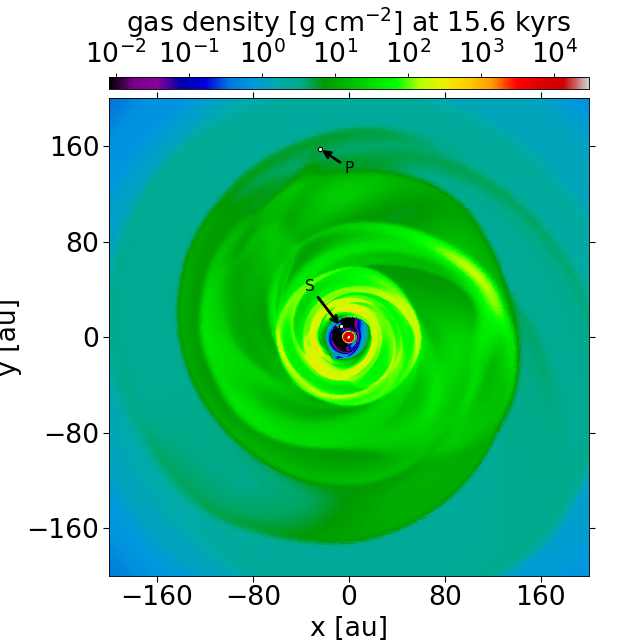}
	\end{minipage}
	\begin{minipage}{0.33\linewidth}
		\centering
		\includegraphics[width=1\linewidth]{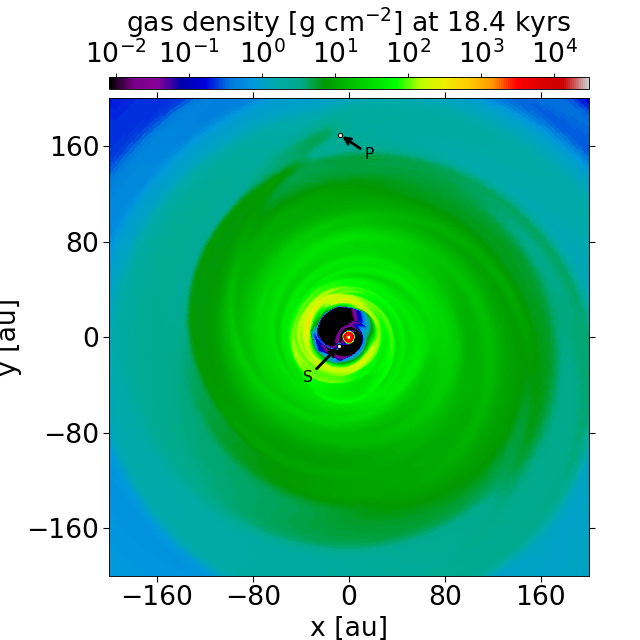}
	\end{minipage}
        \caption{Gas surface density snapshots of the Rp75\_Mp1.0 simulation at different times. The inserted objects are marked by white dots with black edges; `S' and `P' denote the secondary and the planet, respectively. Panels: $t=14.4$~kyr, the encounter between the planet and the secondary; $t=15.6$~kyr, when the secondary has opened a wide gap and the disc exhibits strong dynamical perturbations; and $t=18.4$~kyr, when the disc has developed a significant eccentricity.}

    \label{fig:Rp75_Mp1.0}
\end{figure*}

\begin{figure}
    \centering
    \includegraphics[width=0.45\textwidth]{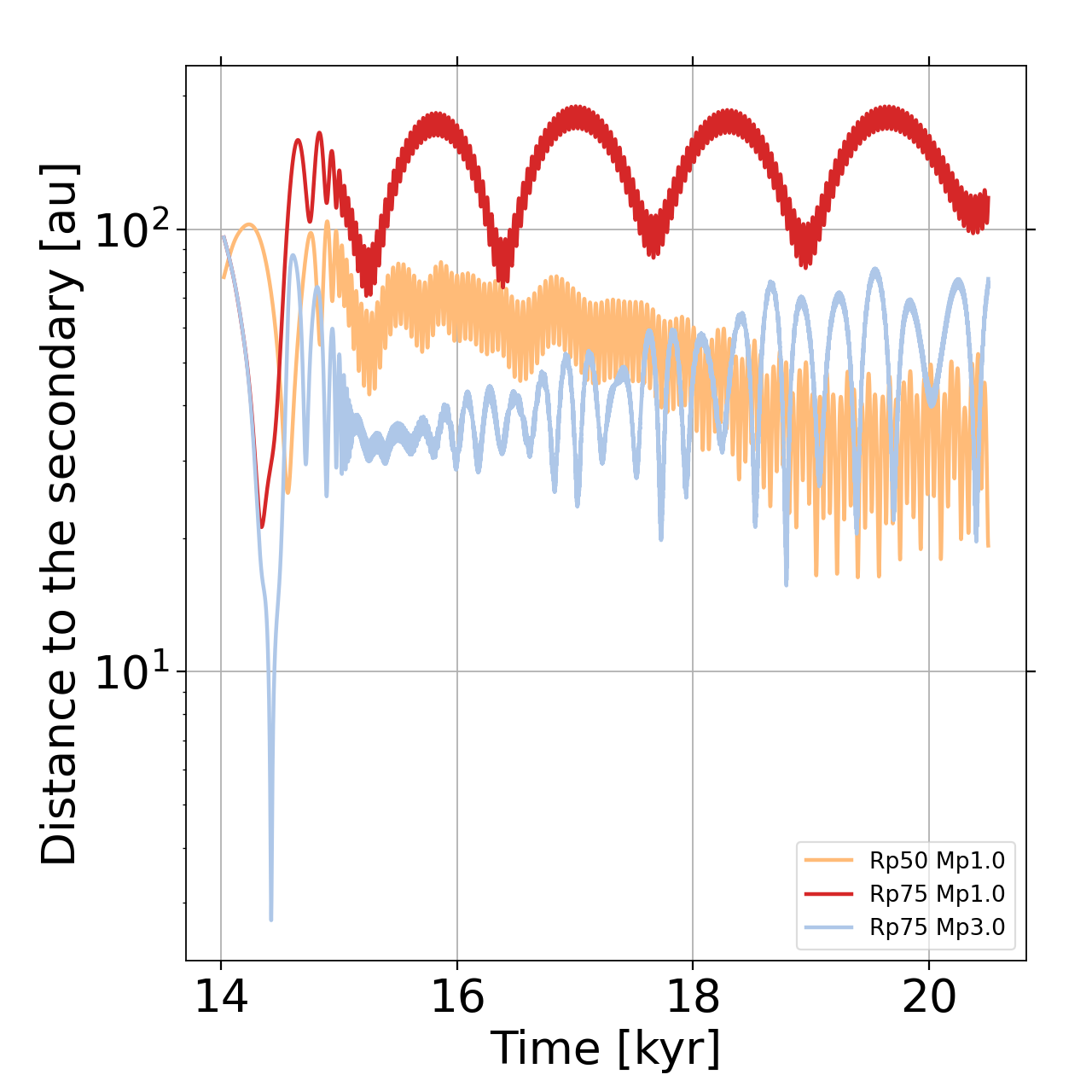}    
    \caption{Time evolution of the separation between secondary and planet in three cases. }
    \label{fig:r0}
\end{figure}

For the Rp50\_Mp1.0 case, the gravitational perturbation from the secondary is relatively weak during the early stages of the evolution. As shown in Fig.~\ref{fig:torque_mr} and Fig.~\ref{fig:r0}, the planet-secondary interaction is not particularly close, resulting in a moderate change of the planet's orbit. The disc torque acting on the planet fluctuates between positive and negative values, resulting in an oscillatory evolution of planet-star separation. Between $t=16$ and $19~\mathrm{kyr}$, however, the time-averaged disc torque is predominantly negative, pulling the planet closer to the secondary. Note that when the planet nears the boundary of the empirically established \cite{Holman_Wiegert_99} instability zone, the secondary's torque grows in strength significantly, exceeding the disc torque in magnitude. The evolution of the two torques at this point is strongly coupled and non-linear. The secondary opens a wide gap in the disc, with width comparable in size to the instability region predicted by \citet{Holman_Wiegert_99}. There is a positive surface density gradient at the gap outer edge (see the right panel in Fig.~\ref{fig:1d-disc}), which could trap a planet there, as expected \citet{Masset06_planet_traps,PierensNelson08}. However, our self-gravitating discs are quite turbulent and time-dependent. The evolving disc structure at the outer gap edge forces the planet-secondary separation to evolve with time, as seen in both Figs \ref{fig:torque_mr} and \ref{fig:r0}. The simulations would need to be continued until the disc dissipation to clarify whether the planet settles onto a stable P-type orbit or is eventually ejected. 

%This outcome appears to be favoured for relatively massive planets that are initially located close to the secondary. In such cases, the planet can remain trapped near the stability boundary for an extended period and may subsequently migrate inward together with the secondary. 

In the Rp75\_Mp1.0 case, the disc torque acting on the planet is generally weaker, but exhibits two prominent positive spikes associated with interactions with the secondary and the subsequent development of spiral structure in the disc. As shown by the disc density snapshots in Fig.~\ref{fig:Rp75_Mp1.0}, the first spike, occurring at $t\approx14.4$~kyr, coincides with a close encounter with the secondary. Besides the strong direct torque from the secondary, a pronounced asymmetry in the gas density develops in the planet's horseshoe region during its encounter. This density contrast generates a strong dynamical corotation torque (Fig.~\ref{fig:torque_mr}), which could contribute to the outward scattering of the planet and the rise of its eccentricity. By $t=15.6$~kyr, the secondary has opened a deep gap and launched prominent spiral density waves, which strongly perturb the surrounding disc and give rise to highly variable torque behaviour. The second positive torque episode is associated with these spiral structures and their interaction with the planet. By $t=18.4$~kyr, the disc has developed a significant eccentricity and the planet has settled onto a wide circumbinary P-type orbit. This evolutionary pathway appears to be more common for lower-mass planets initially located at larger separations from the secondary. In such systems, the planet is more likely to be affected by perturbations from both the secondary and the spiral density waves. As a result, the planet can remain on a wide circumbinary orbit with high eccentricity for an extended period. 

For the Rp75\_Mp3.0 case, a close encounter between the planet and the secondary occurs at $t\approx14.4$~kyr, producing a strong torque impulses (interestingly, from both the secondary and the disc) that scatter the planet into the inner disc. Although such events are rare in our suite of simulations, planets can be scattered into the inner disc, and survive as S-type planets there if they continue to migrate inwards rapidly to avoid further interactions with the secondary. This scenario is most likely for massive planets (that migrate inwards rapidly). It also highlights that the migration of the secondary itself can be noticeably modified by the dynamical back-reaction of a massive planet during close encounters. 

In summary, the presence of the secondary can substantially modify both the disc structure and the migration history of the planet. Planetary migration may be slowed or even halted due to the combined effects of the secondary's gravitational perturbations and the disc torques. In some cases, the planet becomes trapped near the edge of the dynamically unstable region carved by the secondary, while in others it undergoes a more complex evolution driven by violent encounters with the secondary and spiral density waves. Further investigations covering a wider parameter space and longer integration times will be required to fully characterise circumbinary planet formation and migration in these systems.

\bsp	% typesetting comment
\label{lastpage}

\end{document}